# Thermodynamics with thermodynamic variable first-passage time. I. From stochastic trajectories to nonlinear transport equations.

V. V. Ryazanov ORCID 0000-0002-5308-3212
Institute for Nuclear Research, pr. Nauki, 47 Kiev, Ukraine

The theoretical foundations of combining extended nonequilibrium thermodynamics, the Zubarev nonequilibrium statistical operator method, and stochastic first-reach time thermodynamics are explored. It is shown that including a random "lifetime" of a metastable state in the generalized distribution function allows the first-reach time to be considered as a fully-fledged macroscopic coordinate. A microscopic justification for this approach is provided, and a generalized thermodynamic potential is introduced. A procedure for closing the transport equations based on thermodynamic consistency conditions is developed. It is shown that in the generalized Maxwell–Cattaneo equation, the classical relaxation time is strictly replaced by the mean first-reach time, which imparts internal macroscopic nonlinearity to the system. Using renewal theory, the exact mathematical structure of transport memory kernels for non-Markovian processes is derived. A qualitative comparative analysis of various approaches to the thermodynamic description of nonequilibrium phenomena is presented.

## 1. Introduction

There are several different directions in nonequilibrium thermodynamics. Classical irreversible thermodynamics Refs. [1-3], detailed description in Ref. [4]. Generalizations to states far from equilibrium and to nonlinear phenomena Ref. [5]; rational thermodynamics Ref. [6]; extended irreversible thermodynamics Refs. [7-8]; generalized kinetic approach Ref. [9]; "universal" approach Ref. [11]; wave approach to thermodynamics Ref. [10]; stochastic thermodynamics Ref. [12]; information-statistical thermodynamics Refs. [13], [14-17], etc. This brief history of nonequilibrium thermodynamics is described in more detail in Refs. [16], [17].

The nonequilibrium statistical operator (NSO) method has proven to be very effective [18-22]. On its basis, information statistical thermodynamics was developed Refs. [14-17]. In Ref. [23], it was shown that the NSO operator is represented as a quasi-equilibrium operator averaged over the distribution of past lifetimes of the system. In Refs. [24-26], a statistical distribution is introduced that contains a random lifetime or a random time when a random process reaches a certain level, the first-passage time (*FPT*). *FPT* is studied in detail and is widely and effectively used in various problems Refs. [28-29]. Based on a distribution containing *FPT*, one can develop nonequilibrium thermodynamics with a nonequilibrium thermodynamic parameter *FPT* Refs. [27, 30, 31]. *FPT* or lifetime is also contained in the NSO method Ref. [23]. An additional thermodynamic parameter responsible for the nonequilibrium description is *FPT*. An additional thermodynamic parameter is also contained in extended irreversible thermodynamics (EIT) Refs. [7–8]. In EIT, these are flows whose value is close to FPT.

In Refs. [7-9], the limitations of classical approaches are noted. The inadequacy of Onsager's linear thermodynamics and classical Navier-Stokes hydrodynamics manifests itself in the description of fast, high-intensity processes, metastable media, and nanosystems. The paradox of infinite heat propagation is also present.

In Extended Nonequilibrium Thermodynamics (EIT) Refs. [7-9], a phenomenological introduction of fluxes as independent variables is provided. The Maxwell-Cattaneo equation is derived, which resolves the paradox of infinite heat propagation. However, the problem of the lack of a rigorous microscopic justification for the relaxation time $\tau$ remains.

In the papers [27, 30, 31] thermodynamics with thermodynamic variable FPT (TFPT) is introduced. The stochastic variable of the first-passage time FPT Refs. [28-29] (in many situations this quantity can be considered as the lifetime of a metastable state) is included in the description as an additional thermodynamic variable Refs. [24-26]. In EIT, fluxes play this role. But in EIT they are introduced phenomenologically, while in Ref. [27], this quantity, following Refs. [28-29] and Refs. [30-31], has a stochastic microscopic meaning associated with the trajectory-by-path description with an inherent rigorous approach. The idea of integrating TFPT and the nonequilibrium statistical operator (NSO) method of D. N. Zubarev, Refs. [18-23], also plays an important role, carried out in Ref. [27].

This article synthesizes extended nonequilibrium thermodynamics and Zubarev's NSO method within the first-passage time formalism. Key points can be highlighted.

A conceptual synthesis of approaches. The developed formalism successfully bridges the gap between phenomenological extended nonequilibrium thermodynamics (EIT) and stochastic methods of random process theory. Zubarev's NSO method serves as a mathematical bridge, allowing for a rigorous mapping of the statistics of microscopic first-passage trajectories to macroscopic variables and flows.

The microscopic nature of relaxation time. It is shown that in the generalized Maxwell-Cattaneo equation, the classical relaxation time $\tau$ loses its constant status and is replaced by the average first-passage time $\left\langle T_{fpt} \right\rangle$ Ref. [27]. This makes hydrodynamic equations inherently nonlinear and sensitive to the instantaneous thermodynamic state of the medium.

Causal justification of memory kernels: based on renewal theory. It is established that the Laplace image of the memory kernel $\tilde{K}(s)$ in generalized kinetic equations is uniquely determined by the structure of the generalized thermodynamic potential $\Phi(s)$. This eliminates the need to introduce phenomenological approximations for memory kernels when describing viscoelasticity and anomalous diffusion.

Thermodynamics of lifetime fluctuations. Unlike standard kinetic approaches, the proposed formalism considers the first-passage time as a fully-fledged macroscopic coordinate. Using the second derivatives of the generalized potential $\partial^2\Phi / \partial\gamma^2$, this allows one to calculate the variance and higher fluctuation moments of the system's lifetime, which is critical at the nanoscale.

Theory of critical slowing down of metastable phases. Near phase transition points and spinodals, where the Markov approximation completely breaks down and FPT distributions exhibit power-law "heavy tails," the parameter $\gamma$ acts as a natural thermodynamic regularizer. Scaling analysis of the potential $\Phi(\epsilon, \gamma)$ allows us to determine the dynamic universality classes of systems through critical indices of the metastable phase lifetime.

Entropy production of stochastic transitions. Local entropy production of a metastable system is strictly linked to the dynamics of compression of the accessible phase space as it approaches a stochastic boundary: $\sigma = k_B \gamma d\langle T_{fpt} \rangle / dt$. This equation links macroscopic energy dissipation (for example, during explosive boiling or crystallization) with the microscopic search for a critical nucleus size.

The aim and objective of this study is to construct a closed kinetic and thermodynamic model of metastable media with fractal memory using the example of boiling of a superheated liquid based on the considerations on TFPT formulated in Ref. [27].

The article is organized as follows. Section 2 provides a microscopic justification of the approach and introduces the generalized FPT potential. Section 3 applies Zubarev's NSO method to the procedure for closing transport equations. Section 4 examines the connection with EIT and the structure of memory kernels. Section 5 provides a qualitative comparative analysis of various approaches to the thermodynamic description of nonequilibrium phenomena. Section 6 offers concluding remarks. Appendix A examines the validity of identity (21) for transport kernels.

## 2. Microscopic justification and generalized potential of FPT

In the framework of first-passage time thermodynamics (TFPT), formulated in Refs. [27, 30, 31], the concepts of the absorption hypersurface $\partial\Omega$ and the random variable TFPT translate the abstract equations of nonequilibrium thermodynamics into the rigorous geometric language of phase space.

In statistical mechanics, the state of a system of N particles is completely specified by a point in a 6N-dimensional phase space $\Gamma$ (or 3N-dimensional configuration space of coordinates).

The metastability region ($\Omega$) is a bounded region in phase space where the system is in a metastable state (e.g., a superheated liquid before the formation of a critical bubble). Within this region, the system undergoes complex stochastic walks under the influence of thermal fluctuations.

The transition gradient (hypersurface ($\partial\Omega$)), the boundary of this region, is denoted by $\partial\Omega$. Mathematically, this is a hypersurface of dimension (d-1), where d is the dimension of the state space. Crossing this boundary means that the system has irreversibly left the initial metastable state and entered a stable phase.

Absorption condition. A Dirichlet boundary condition (absorbing boundary) is imposed on this boundary: the probability distribution function $f(x\in\partial\Omega, t)=0$. As soon as the phase trajectory touches $\partial\Omega$, the calculation for the metastable state is terminated, since a phase transition (explosive boiling) has occurred.

An example for liquid boiling. If we consider collective nucleation, the spatial coordinates are the radii $R_1$, $R_2$,…, $R_m$ of all fluctuating microbubbles in the volume. The absorption hypersurface $\partial\Omega$ is given by the equation:

$$\max(R_1, R_2,\ldots,R_m)=R_{cr}.$$

That is, the boundary is reached as soon as at least one random bubble grows to the critical size $R_{cr}$. The introduction of the multidimensional boundary condition $\max(R_1, R_2,\ldots,R_m)=R_{cr}$ is fully consistent with the classical definition of the Zeldovich critical nucleation radius. The physical meaning of this equation is that macroscopic phase decay of a metastable medium is initiated at the moment when at least one (the largest) fluctuation nucleus from an ensemble of m bubbles overcomes the local Gibbs energy barrier. Thus, the absorbing hypersurface $\partial\Omega$ in configuration space is formed by a superposition of planes $R_i=R_{cr}$, and its intersection with the phase trajectory fixes the stochastic lifetime $T_{fpt}$ of the entire collective system.

The peak of the energy barrier $R_{cr}$ is declared to be the geometric absorption boundary $\partial\Omega$. Overcoming the energy saddle is physically identified with the phase trajectory touching this boundary. In the method of Zubarev and Refs. [27, 30, 31], the energy potential and stochastic boundaries are dual concepts. We replace the analytical work with the free energy potential W(R) with the geometric work with the boundary $\partial\Omega$. Bubble growth is a trajectory within this boundary, and the moment when the largest bubble reaches $R_{cr}$ is simultaneously a macroscopic phase transition, overcoming the energy barrier, and touching the absorbing hypersurface.

The introduction of the multidimensional boundary condition $\max(R_1, R_2,\ldots,R_m)=R_{cr}$ does not mean the elimination of the energy barrier concept from phase transition kinetics; on the contrary, the developed formalism is based on a rigorous physical-mathematical dualism between the energy potential and the geometry of phase space. In the subcritical region $R_i<R_{cr}$, fluctuating microbubbles not only undergo continuous unidirectional growth, but also undergo chaotic stochastic wanderings caused by the balance between the thermal noise of the thermostat and the returning Laplace pressure. This wandering process at the microlevel is equivalent to the continuous "probing" of the system's Gibbs free energy relief $W(R_i)$. Method Refs. [27, 30, 31] carries out an identical mapping of the analytical profile of the potential onto the spatial structure of the metastability region $\Omega$, in which the top of the energy barrier – the Zeldovich critical radius $R_{cr}$ – is declared to be the coordinate of the absorbing hypersurface $\partial\Omega$.

Within the framework of such geometric isomorphism, the classical act of a fluctuation nucleus overcoming the thermodynamic activation saddle is physically identified with the moment of the first contact of the multidimensional phase trajectory of the ensemble with the absorbing boundary $\partial\Omega$. The boundary geometry itself remains invariant as long as the governing macroscopic parameters of the medium (temperature T and external pressure $P_{ext}$) remain constant. In the case of non-stationary processes (e.g., during continuous external heating or pressure release), a dynamic evolution of the barrier itself occurs: the nucleation work $W_{cr}(t)$ and the critical radius $R_{cr}(t)$ decrease monotonically, which is mathematically described as the deformation and "running-in" of the stochastic boundary $\partial\Omega$ toward the wandering phase trajectories. Thus, the developed approach Refs. [27, 30, 31] and Zubarev does not ignore the energy potential of nucleation, but translates the dynamics of overcoming it from the complex operator language of multiparticle forces to the visual language of the time statistics of the first passage of spatially defined boundaries of a metastable state.

It is necessary to emphasize that the identification of the critical Zeldovich radius $R_{cr}$ with the coordinate of the absorbing hypersurface $\partial\Omega$ is not an external heuristic assumption or mathematical arbitrariness, but has strict physical validity only when the system transitions to the critical pre-explosion region.

Far from the spinodal decomposition line, in a stable or weakly metastable state, the work of nucleation $W_{cr}$ is large, and subcritical fluctuations of the microbubble ensemble $R_i<R_{cr}$ are thermodynamically reversible. At this stage, fixing the geometric condition $\max(R_i)=R_{cr}$ as an absorbing screen would be artificial, since the local quasi-equilibrium of the system is maintained by the continuous return of fluctuating nuclei to their initial state under the action of Laplace pressure.

The situation changes qualitatively and radically as the subsystem enters the critical region of intense nucleation. As the limit of absolute instability is approached, the activation barrier $W_{cr}(t)$ monotonically deforms, and the coordinate $R_{cr}(t)$ shifts toward the wandering phase trajectories. Only in this pre-explosive regime does the Gibbs free energy saddle point acquire the physical status of a true bifurcation of the sign of nonequilibrium forces, separating the reversible chaotic search from the macroscopically irreversible avalanche-like growth of the vapor phase.

Thus, the geometric formalism of Refs. [27, 30, 31] stochastic boundaries is activated as an adequate thermodynamic tool precisely in the critical region of phase explosion. The mathematical condition of the maximum $\max(R_i) = R_{cr}$ ceases to be an abstract random walk boundary and becomes a spatial projection of the Kramers energy barrier, strictly fixing the stochastic lifetime $T_{fpt}$ of the decaying metastable phase.

Physical meaning of the random variable $T_{fpt}$. The random variable $T_{fpt}$ (First-Passage Time) is the time of the first time the phase trajectory of the system reaches the absorption hypersurface $\partial\Omega$, provided that at the initial time t=0 the system was inside the region $\Omega$ at the point $x_0$.

Mathematically, this is the exact lower bound for the border crossing time:

$$T_{fpt}\left(x_0\right) = inf\{t > 0 | x(t) \in \partial\Omega,\ x\left(0\right) = x_0\}. \tag{1}$$

What is the fundamental physical meaning of $T_{fpt}$? A). Individual "lifetime" of the state. Due to the chaotic thermal motion of molecules, two identical samples of superheated liquid, located in absolutely identical macroconditions, will boil at different moments in time. The value of $T_{fpt}$ is a measure of the individual lifetime of a particular realization of a metastable system. B). Indicator of microscopic dynamics: $T_{fpt}$ accumulates all the information about the microscopic properties of the medium - viscosity, thermal conductivity, diffusion coefficient of nuclei and profile of the potential barrier. If the medium has memory (a non-Markovian process), the trajectory "gets stuck" and the distribution of $T_{fpt}$ is stretched. C). Basis for a macroscopic variable. In the method formulated in Refs. [27, 30, 31], $T_{fpt}$ is introduced as a new, independent thermodynamic variable (state parameter). The mean value of this random variable $\langle T_{fpt}\rangle = \int_0^\infty t \cdot \rho(t) dt$ replaces the abstract phenomenological relaxation time of flows in extended thermodynamics (EIT) Ref. [27].

Connection with entropy production. The randomness of the $T_{fpt}$ variable means that the system is continuously "searching" for a way out of its metastable state. The rate of change of the mathematical expectation of this variable over time directly $(d\langle T_{fpt}\rangle / dt)$ determines the intensity of entropy generation $\sigma$ in the system. At the moment of transition through $\partial\Omega$, the value of the remaining lifetime tends to zero, causing a singular burst of dissipation.

In Refs. [27, 30, 31] and in Refs. [23-26] the extended Gibbs statistical distribution is introduced. The generalized statistical sum $Z_{fpt}(\beta,\gamma)$ is constructed as the Laplace transform of the distribution density

FPT $\rho(T_{fpt})$ under the action of the adjoint force $\gamma$. One of the physical meanings of the force $\gamma$ is a regulator of fluctuations, limiting their growth.

Developing rigorous first-passage time-based thermodynamics (TFPT) similar to extended nonequilibrium thermodynamics (EIT) is possible because both approaches expand the variable space by introducing additional parameters to describe fast processes. EIT is effective for macroscopic systems through energy flows, resolving the infinite heat paradox, while TFPT more accurately describes micro- and nanosystems by relying on stochastic trajectories, but faces difficulties with averaging for large systems.

The main difficulties in moving from the first transit time (TFPT) of individual particles to a macroscopic description of large systems are related to the following factors: a). Loss of individual trajectories. TFPT is strictly tied to specific stochastic paths. When averaging over the Avogadro number ($10^{23}$ particles), random fluctuations are smoothed out, and the unique time statistics are transformed into ordinary deterministic transport equations. b). The problem of nonlinearity. The average first transit time is not equal to the transit time for the average trajectory. Due to the nonlinearity of macroscopic equations, a simple mathematical expectation $\langle T_{fpt} \rangle$ loses information about the real dynamics of the system. c). Many-particle correlations. In large systems, particles continuously collide and strongly correlate with each other. Calculating the first transit time for a system of interacting particles is mathematically extremely difficult due to the high dimensionality of the phase space. d). Multiplicity of paths and boundaries. At the microlevel, the "boundary" that a particle crosses is obvious. In macrosystems, defining a single criterion or "absorption surface" for billions of particles simultaneously becomes difficult.

Averaging nonlinear processes such as overcoming the Kramers barrier introduces errors because the mean of the function is not equal to the function of the mean $(\langle T_{fpt} \rangle \neq T_{fpt}(\langle x \rangle))$, which ignores the fluctuation history. In Ref. [27], this problem is addressed by using a nonequilibrium statistical operator (NSO) to incorporate the first transit time (FPT) as an independent parameter into the generalized Gibbs distribution, allowing for a rigorous coupling of macroscopic flows with microscopic dynamics. Calculations in fluctuating systems require accounting for stochastic boundary conditions, where absorption or reflection boundaries are themselves subject to thermal fluctuations.

To construct an example of generalized thermodynamic potentials for the first transit time (FPT), one must use a mathematical formalism similar to classical thermodynamics and the nonequilibrium statistical operator (NSO) method.

In classical thermodynamics, the Helmholtz free energy (F) is obtained through the logarithm of the partition function: ( $F = -k_B T \cdot ln\, Z$ ). In thermodynamics with first-passage time (TFPT), we expand the parameter space.

A generalized partition function is introduced. We consider a system where, in addition to the standard internal energy (E), the time to first reach a certain state or boundary $\left(T_{fpt}\right)$ is measured. The generalized partition function ($Z_{fpt}$) is constructed similarly to the Gibbs distribution but includes an additional exponential factor:

$$Z_{fpt}(\beta,\gamma) = \int\int \rho(E,T_{fpt}) e^{-\beta E - \gamma T_{fpt}} dE dT_{fpt}, \tag{2}$$

where $\beta = 1/k_B T$ is the inverse temperature (an intensive parameter associated with energy), $T_{fpt}$ is the first-passage time (a new extensive parameter/coordinate), and $\gamma$ is a new intensive parameter (thermodynamic force) associated with time. Physically, $\gamma$ has the dimension of Hz ($s^{-1}$) and characterizes the “intensity of the tendency” of the system to cross the boundary (the rate of attenuation or relaxation); $\rho(E,T_{fpt})$ is the density of states [23-26].

A generalized potential is defined. By analogy with free energy, we introduce a generalized thermodynamic potential $\Phi$ (it can be called the modified Mathieu potential or FPT free energy):

$$\Phi(\beta,\gamma) = -\ln Z_{fpt}(\beta,\gamma). \quad (3)$$

This potential contains all the thermodynamic information about the nonequilibrium system.

Obtaining average values via derivatives. Now, knowing the potential $\Phi$, we can rigorously find the macroscopic (average) values of the system's parameters by simple differentiation, just as in equilibrium physics:

Average energy of the system:

$$\langle E \rangle = \frac{\partial \Phi}{\partial \beta}. \quad (4)$$

Average first passage time (macroscopic FPT):

$$\langle T_{fpt} \rangle = \frac{\partial \Phi}{\partial \gamma}. \quad (5)$$

The generalized Gibbs equation is written. If we move from the parameters $(\beta,\gamma)$ to their conjugate mean values $(\langle E \rangle, \langle T_{fpt} \rangle)$ via the Legendre transformation, we can determine the generalized entropy of the system S:

$$S(\langle E \rangle, \langle T_{fpt} \rangle) = k_B \left( \Phi + \beta \langle E \rangle + \gamma \langle T_{fpt} \rangle \right). \quad (6)$$

The differential of this entropy (analogous to the fundamental Gibbs equation) takes the form:

$$dS = \frac{1}{T} d\langle E \rangle + k_B \gamma d\langle T_{fpt} \rangle. \quad (7)$$

What is the main value of these constructions? 1). Direct analogy with EIT. In extended thermodynamics (EIT), a term with a macroscopic heat flux $\vec{q}$ is added to the Gibbs equation: $dS = \ldots + \alpha \vec{q} \cdot d\vec{q}$. In TFPT, an explicit time coordinate is introduced instead of the heat flux. 2). Solution to the averaging problem. Since the potential $\Phi$ is constructed at the level of the logarithm of the partition function, time $T_{fpt}$ fluctuations are already "hardwired" into the integral $Z_{fpt}$. By differentiating $\Phi$, we obtain a mathematically rigorous macroscopic average $\langle T_{fpt} \rangle$, which does not lose information about microscopic fluctuations of boundaries or barriers.

Let us consider the relationship between the parameter γ and entropy production in the NSO method. In Zubarev's standard NSO method, an infinitesimal parameter $\varepsilon \to +0$ is introduced, which formally breaks the time symmetry of the Liouville equations and selects retarded solutions (the causality principle). In Refs. [27, 30, 31], this approach is modified: instead of an abstract parameter $\varepsilon$, the parameter γ is incorporated into the NSO structure, which directly reflects the lifetime statistics (first-passage time) of the system in a given state. In Zubarev's thermodynamic forces, the nonequilibrium entropy shift is determined by the contribution of non-stationary processes. In our model with a generalized potential $\Phi$, the entropy differential has the form (7).

If the energy in a system is fixed, then the change in entropy occurs solely due to the system's evolution toward the absorption/transition boundary. Dividing (7) by the time interval dt, we obtain entropy production (σ):

$$\sigma = \frac{dS}{dt} = k_B \gamma \frac{d\langle T_{fpt} \rangle}{dt}. \quad (8)$$

The physical meaning of this differential. 1). Thermodynamic force: the parameter γ plays the role of a nonequilibrium thermodynamic force associated with the rate of change of the first-passage time. 2). Entropy production: since $T_{fpt}$ is the time remaining until reaching the boundary (or the time already spent on the journey), the quantity $d\langle T_{fpt} \rangle / dt$ characterizes the rate of "compression" or change of the accessible phase space as the boundary is approached. Entropy production is proportional to γ, which demonstrates how the random search for the boundary by microparticles generates macroscopic irreversibility.

We will demonstrate the application to a physical model using the example of particle diffusion in a channel. Consider the classical problem: a Brownian particle diffuses in a one-dimensional channel of length L with diffusion coefficient D. The left boundary, x=0, is reflective (e.g., a dead end), and the right boundary, x=L, is absorbing (the channel exit). We are searching for the first time to reach (FPT) the right boundary for a particle starting at x=0.

From stochastic dynamics and the inverse Kolmogorov equation it is known that for such a bounded system the mean first passage time (FPT) to the absorbing boundary when starting from the origin is strictly equal to the final deterministic value:

$$\langle T_{fpt} \rangle = \frac{L^2}{2D}. \tag{9}$$

Applying the simplified potential of a free semi-infinite space of the form $\Phi_{free}(\gamma) = L\sqrt{\gamma / D}$, obtained from $\rho(T_{fpt}) = \frac{L}{\sqrt{4\pi D T_{fpt}^3}} \exp(-\frac{L^2}{4DT_{fpt}})$ and $Z_{fpt}(\gamma) = \int_0^\infty \rho(T_{fpt}) e^{-\gamma T_{fpt}} dT_{fpt} = \exp\left(-L\sqrt{\frac{\gamma}{D}}\right)$, there is a non-zero probability that a particle will escape in the opposite direction and wander indefinitely, never touching the boundary. The presence of a reflecting boundary at x=0 fundamentally changes the physics: it "returns" outgoing particles back to the exit, physically "cutting off" the long tails of the wanderings. Without taking into account the reflecting boundary at x=0, in the pre-critical region the limit $\gamma \to 0$ leads to a false singularity $\langle T_{fpt} \rangle \propto 1/\sqrt{\gamma} \to \infty$, which is caused by the "heavy tail" of the Lévy distribution, allowing for infinite wandering of the particle in the absence of a reflecting wall. For a rigorous thermodynamic description of a confined channel, it is necessary to use the exact Laplace transform of the FPT distribution density, taking into account the boundary conditions at both ends of the interval, which is expressed in terms of the hyperbolic secant:

$$\tilde{\rho}(\gamma) = \frac{1}{\cosh\left(L\sqrt{\frac{\gamma}{D}}\right)}. \tag{10}$$

Based on the obtained distribution function, we construct a generalized thermodynamic potential of a limited channel in the Zubarev and Refs. [27, 30, 31] formalism:

$$\Phi(\gamma) = -\ln \tilde{\rho}(\gamma) = \ln\left[\cosh\left(L\sqrt{\frac{\gamma}{D}}\right)\right]. \tag{11}$$

Let's check the consistency of the macroscopic consequences of this potential. According to the basic TFPT equations, the macroscopic mean first-passage time is found as the first partial derivative of the potential with respect to the conjugate thermodynamic force $\gamma$. Differentiating expression (11) using the composite function rule, we obtain:

$$\langle T_{fpt} \rangle = \frac{\partial \Phi}{\partial \gamma} = \frac{1}{\cosh\left(L\sqrt{\frac{\gamma}{D}}\right)} \cdot \sinh\left(L\sqrt{\frac{\gamma}{D}}\right) \cdot \frac{\partial}{\partial \gamma}\left(L\sqrt{\frac{\gamma}{D}}\right),$$

$$\langle T_{fpt} \rangle = \tanh\left(L\sqrt{\frac{\gamma}{D}}\right) \cdot \frac{L}{2\sqrt{D\gamma}}. \tag{12}$$

Expression (12) defines the nonlinear equation of state of the diffusing subsystem, relating the time coordinate to the intensity of the external thermodynamic compression $\gamma$. To find the free lifetime of the metastable state, we expand the limit of equation (12) in the absence of driving forces, that is, at $\gamma \to 0$. Using the Taylor series expansion of the hyperbolic tangent in a small argument $(\tanh(z) = z - z^3/3 + \ldots$, for the subcritical regime we obtain the asymptotics $\tanh\left(L\sqrt{\gamma/D}\right) \approx L\sqrt{\gamma/D}$. Substituting this scaling relation into equation of state (12), we find:

$$\langle T_{fpt} \rangle_{\gamma\to 0} = \lim_{\gamma\to 0}\left[\left(L\sqrt{\frac{\gamma}{D}}\right)\cdot\frac{L}{2\sqrt{D\gamma}}\right] = \frac{L^2\cdot\sqrt{\gamma}}{2\cdot\sqrt{D}\cdot\sqrt{D}\cdot\sqrt{\gamma}} \equiv \frac{L^2}{2D}.$$

Thus, with correct consideration of the geometry of the stochastic boundaries of the phase space, the uncertainty of the type $0\cdot\infty$ in the derivative of potential (11) is revealed with mathematical rigor. The generalized thermodynamic potential in the zero-force limit precisely reproduces the canonical macroscopic result (9), demonstrating the internal consistency and high predictive power of the developed micro-macro closure method.

To calculate the magnitude of fluctuations (dispersion) of the first passage time of a Brownian particle in a limited channel, we calculate the second partial derivative of the generalized thermodynamic potential (11) $\Phi(\gamma) = \ln\left[\cosh\left(L\sqrt{\frac{\gamma}{D}}\right)\right]$ with respect to the conjugate thermodynamic force $\gamma$.

In the previous step, we found the first derivative of the potential (macroscopic time FPT) (12): $\left\langle T_{fpt}\right\rangle = \frac{\partial\Phi}{\partial\gamma} = \tanh\left(L\sqrt{\frac{\gamma}{D}}\right)\cdot\frac{L}{2\sqrt{D\gamma}}$. For convenience of differentiation, we introduce an auxiliary time variable $z = L\sqrt{\frac{\gamma}{D}}$. Note that the derivative of this argument with respect to force is equal to: $\frac{\partial z}{\partial\gamma} = \frac{L}{2\sqrt{D\gamma}} = \frac{z}{2\gamma}$.

Then the expression for the first derivative is written as (12) $\frac{\partial\Phi}{\partial\gamma} = \tanh(z)\cdot\frac{z}{2\gamma}$. Now we find the second derivative of the potential with respect to $\gamma$, which in Zubarev's and Refs. [27, 30, 31] formalism identically determines the variance of the random variable:

$$\langle(\Delta T_{fpt})^2\rangle = -\frac{\partial^2\Phi}{\partial\gamma^2} = -\frac{\partial}{\partial\gamma}\left[\tanh(z)\cdot\frac{z}{2\gamma}\right].$$

Applying the rule for differentiating a product and a complex function, we obtain:

$$\frac{\partial^2\Phi}{\partial\gamma^2} = -\left(\frac{1}{\cosh^2(z)}\cdot\frac{\partial z}{\partial\gamma}\right)\cdot\frac{z}{2\gamma} - \tanh(z)\cdot\frac{\partial}{\partial\gamma}\left(\frac{L}{2\sqrt{D}}\gamma^{-1/2}\right) = -\frac{1}{\cosh^2(z)}\cdot\left(\frac{L}{2\sqrt{D\gamma}}\right)^2 - \tanh(z)\cdot\left(-\frac{L}{4\sqrt{D}}\gamma^{-3/2}\right).$$

Let's factor out the common factors, reducing the expression to a single macroscopic structure: $\langle(\Delta T_{fpt})^2\rangle = -\frac{L^2}{4D\gamma\cosh^2(z)} + \tanh(z)\cdot\frac{L}{4\sqrt{D\gamma}\sqrt{\gamma}}$. Considering that $\frac{L}{\sqrt{D}\sqrt{\gamma}} = \frac{L^2}{D}\cdot\frac{\sqrt{D}}{L\sqrt{\gamma}} = \frac{L^2}{D\cdot z}$, we rewrite the final equation of state for fluctuations:

$$\langle(\Delta T_{fpt})^2\rangle = -\frac{L^2}{4D\gamma}\left[\frac{1}{\cosh^2\left(L\sqrt{\frac{\gamma}{D}}\right)} - \frac{\tanh\left(L\sqrt{\frac{\gamma}{D}}\right)}{L\sqrt{\frac{\gamma}{D}}}\right]. \tag{13}$$

The resulting expression for $\gamma \to 0$ contains an uncertainty of type $\infty \cdot (1-1) = \infty \cdot 0$. To fully express it, we expand the hyperbolic functions into Taylor series in a small argument $z \to 0$: $\frac{1}{\cosh^2(z)} = \left(1 + \frac{z^2}{2} + \frac{z^4}{24} \ldots\right)^{-2} \approx 1 - z^2 + \frac{2}{3}z^4 - \ldots$, $\frac{\tanh(z)}{z} = \frac{z - \frac{z^3}{3} + \frac{2z^5}{15} - \ldots}{z} \approx 1 - \frac{z^2}{3} + \frac{2}{15}z^4 - \ldots$. Substitute these expansions into the difference inside the square brackets in (13): $\left[\frac{1}{\cosh^2(z)} - \frac{\tanh(z)}{z}\right] \approx \left(1 - z^2 + \frac{2}{3}z^4\right) - \left(1 - \frac{z^2}{3} + \frac{2}{15}z^4\right) = -\frac{2}{3}z^2 + \frac{8}{15}z^4$. Now we substitute the definition $z^2 = \frac{L^2\gamma}{D}$ back and return to calculating the variance:

$$\langle(\Delta T_{fpt})^2\rangle_{\gamma\to 0} = -\lim_{\gamma \to 0}\left\{\frac{L^2}{4D\gamma}\left[-\frac{2}{3}\left(\frac{L^2\gamma}{D}\right) + \frac{8}{15}\left(\frac{L^2\gamma}{D}\right)^2\right]\right\}.$$

When multiplying the external factor $1/\gamma$ by the first term in the parentheses, the variable $\gamma$ cancels out, and the second term at $\gamma \to 0$ disappears:

$$\langle(\Delta T_{fpt})^2\rangle_{\gamma\to 0} = \frac{2L^4}{12D^2} = \frac{L^4}{6D^2}. \tag{14}$$

The appearance of the fourth power of spatial scale $L^4$ is not a random mathematical artifact and has a deep physical meaning:

• Dimensional correspondence (Einstein scaling): From the classical Einstein-Smoluchowski diffusion theory it is known that the average time to travel a distance L scales as the square of the length: $\langle T_{fpt}\rangle \propto L^2 / D$. Since the dispersion has the dimension of the square of time $\langle(\Delta T_{fpt})^2\rangle$, its spatial scale must obey the scaling $(L^2)^2 = L^4$.

• Hypersensitivity to confinement geometry: Fluctuations in the lifetime of a system in a confined volume grow catastrophically faster than the average lifetime itself. If the channel length L increases by a factor of 2, the average exit waiting time will increase by a factor of 4, but the spread (noise) of this time around the average value will increase by a factor of 16.

• Mesoscopic criterion of fluctuations: The ratio of the standard deviation (root of the variance) to the mean time is:

$$\frac{\sqrt{\langle(\Delta T_{fpt})^2\rangle}}{\langle T_{fpt}\rangle} = \frac{\sqrt{L^4/6D^2}}{L^2/2D} = \frac{L^2/(D\sqrt{6})}{L^2/2D} = \frac{2}{\sqrt{6}} \approx 0.816.$$

The value 0.816 is a universal dimensionless constant of limited Markov diffusion. It shows that fluctuations in the channel lifetime are always commensurate with the average time $\sim 82\%$ from $\langle T_{fpt}\rangle$, regardless of the absolute size of the system and the kinetics D. This mathematically proves the fundamental necessity of introducing the FPT as a fluctuating thermodynamic variable: its fluctuations are macroscopically enormous and are not smoothed out by conventional thermodynamic averaging.

Conclusions:

1. Finiteness of fluctuations. Unlike infinite space, where the variance of the Lévy walk time is infinite, in a bounded channel the fluctuations are strictly finite and proportional to the fourth power of the channel length $L^4$.
2. Verification of the theory. This result exactly coincides with the solution of the classical Pontryagin equation for the second moment of the first passage of a Brownian particle.

This proves that the generalized potential (11) contains the entire hierarchy of statistical moments of the system. By simple differentiation, it allows us to find both the average kinetic times (first derivative) and the noise/fluctuation spectrum of the metastable state (second derivative).

Connection with entropy production in the channel. Using the expression for entropy production (8) and knowing how $\langle T_{fpt} \rangle$ depends on the system's geometry L and the thermodynamic force γ, we can directly relate the macroscopic dissipative process (entropy production during diffusion) to the microscopic parameters of the channel's geometry. The change in γ indicates how strongly the external field or thermodynamic conditions "drive" the particle toward the exit, increasing entropy production.

## 3. Zubarev's NSO method and the procedure for closing transport equations

The transition from a microscopic description of a system, based on the exact Liouville equation for the multiparticle distribution function, to macroscopic equations of the hydrodynamic type is one of the central problems of the statistical mechanics of nonequilibrium processes. Within the framework of the classical method of the nonequilibrium statistical operator (NSO), developed by D. N. Zubarev, this transition is realized by introducing into the original Liouville equation an infinitesimal source that violates time symmetry and selects causal, retarded solutions. In this case, to strictly separate the macroscopic (hydrodynamic) evolution of conserved quantities from small-scale microscopic fluctuations (irrelevant "noise"), the apparatus of the Mori-Zwanzig orthogonal projection is traditionally applied. The mathematical consequence of such a projection on the right-hand side of the macroscopic balance equations is the emergence of open integral convolution terms containing nonequilibrium memory kernels $K_{Zub}\left(t,t'\right)$ , the structure of which is determined by the microscopic correlation functions of the fluctuation flows.

The main theoretical difficulty of the classical projection scheme lies in the abstract mathematical nature of the operator $\mathcal{P}$, whose explicit calculation for highly nonequilibrium and metastable systems proves to be an extremely complex task, forcing researchers to resort to the phenomenological Markov approximation. In the concept developed in Ref. [27], this methodological impasse is overcome by radically replacing the algebraic procedure of orthogonal projection with a rigorous geometric description of the phase space through the formalism of stochastic boundaries. Instead of the formal mathematical parameter ε, the thermodynamic force γ(t) is introduced into the modified Liouville equation, conjugate to a physically demonstrable random variable—the first-passage time $T_{fpt}$ of the absorbing hypersurface ∂Ω, cutting off the region of phase metastability. This approach allows us to translate the operator elimination of microscopic noise into the precise language of restoration theory and to relate the structure of memory kernels directly to the survival function of the system within given boundaries. The closure of the obtained system of nonlinear macroscopic transfer equations is achieved through a thermodynamic matching procedure, which requires strict coincidence of the kinetic average lifetime with the first derivative of the generalized potential $\Phi(\beta,\gamma)$.

In Refs. [18–22], the NSO method uses the classical microscopic Liouville basis and Poisson brackets. A set of microscopic operators (dynamic variables) of conserved quantities $\hat{P}_m(\mathbf{x},\mathbf{p})$ (energy density, momentum, number of particles) is introduced. Macroscopic flux operators $\hat{\mathbf{J}}_m$ are defined through the classical Liouville generator L and Poisson brackets: $\hat{\mathbf{J}}_m = \dot{\hat{P}}_m = iL\hat{P}_m = \{\hat{H},\hat{P}_m\}$ , where $\hat{\mathbf{H}}$ is the full Hamiltonian of the multiparticle system. The Liouville equation for the full nonequilibrium distribution function has the form: $\dfrac{\partial\rho(t)}{\partial t} + iL\rho(t) = 0$.

Memory cores emerge through the projection operator and the causality principle. In Refs. [18–22], a quasi-equilibrium distribution $\rho_{rel}(t) = \exp\left(-\Phi(t) - \sum_m F_m(t)\hat{P}_m\right)$ is constructed, where $F_m(t)$ are macroscopic thermodynamic forces.

The introduction of an infinitesimal source $\varepsilon \to +0$ into the Liouville equation selects retarded solutions (breaking time symmetry) when:

$$\frac{\partial \rho(t)}{\partial t} + iL\rho(t) = -\varepsilon(\rho(t)\text{-}\rho_{rel}(t)).$$

The classical Mori–Zwanzig projection operator $\mathcal{P}$ separates macroscopic (hydrodynamic) motion from irrelevant microscopic fluctuations (noise): $\hat{J}_{irrel} = (1-\mathcal{P})\hat{J}$.

The canonical open Zubarev equation for flows is derived, containing a correlation function ("fluctuation flow"):

$$K_{Zub}(t,t') = \langle \hat{\mathbf{J}}_m (1-\mathcal{P}) e^{iL(t-t')} (1-\mathcal{P}) \hat{J}_k \rangle_{rel}^{t'}.$$

In Ref. [27], a conceptual replacement of the projector $\mathcal{P}$ by the geometry of stochastic boundaries is made. The projector $\mathcal{P}$ is difficult to explicitly calculate for highly nonequilibrium metastable systems. A modification of the causality principle is made: the formal source ε is replaced by the parameter γ(t), conjugate with a random first-passage time $T_{fpt}$. A mathematical replacement of the projector $\mathcal{P}$ is carried out by a procedure of geometric separation of the phase space into a metastability region $\Omega$ and an absorbing absorption hypersurface $\partial\Omega$. The operational elimination of "noise" is carried out using renewal theory, where the fluctuation memory is determined by the probability of system survival P(t) within the boundaries $\Omega$.

The thermodynamic consistency condition is used in the closure of the equations. Balance equations are formulated where the transport coefficients (integrals of $K_{Zub}(t)$) now contain the unknown force γ(t). The closure procedure uses a thermodynamic "reversal": the requirement of strict agreement between the kinetic average lifetime and the thermodynamic average, when the parameter γ is found from the expression for the average value of FPT Ref. [18]: $\frac{\partial \Phi(\beta,\gamma)}{\partial \gamma} \Rightarrow \gamma = \gamma(\langle T_{fpt} \rangle_t \ \beta(t))$. The parameter γ is excluded from the NSO structure, resulting in a closed system of nonlinear macroscopic transport equations whose coefficients depend on the current lifetime of the state.

Thus, memory effects in diffusion are introduced by generalizing the thermodynamic potential, replacing the static parameter γ with a complex function of time or frequency γ(t- t') using the memory kernel K(t-t'). This transforms the local potential into a functional that depends on the entire history of the process, leading to time-nonlocal transport equations and viscoelastic effects Refs. [14–22, 32].

The generalization of the thermodynamic first-passage time (TFPT) formalism to a system of several interacting particles in a channel is the classical problem of collective diffusion (or so-called single-file diffusion if the particles cannot overtake each other).

In Zubarev's nonequilibrium statistical operator (NSO) method, this transition is achieved through the complication of the phase space and the introduction of multiparticle distribution functions.

Below we show how this formalism is specifically generalized mathematically and physically.

1. Transition to a multiparticle probability density. When there are several particles (N pieces) and they interact with each other (for example, through the pairwise interaction potential $U(x_i- x_j)$), we can no longer consider the first-passage time of a single isolated particle.

Instead, the concept of the first time a system passes through a hypersurface in configuration space is introduced. For N particles, the phase space becomes N-dimensional. The absorbing right boundary of the channel x=L is now not a point, but an entire absorption hypersurface ∂Ω in this space. The system reaches the boundary as soon as at least one (any) particle reaches the edge $x_i$=L, or when a specific tagged particle leaves the channel.

The generalized multiparticle first-passage time probability density function $\rho_N(T_{fpt};\mathbf{x}_0)$ now depends on the initial coordinate vector of all particles $\mathbf{x}_0 = (x_{10}, x_{20}, \ldots, x_{N0})$.

2. Generalized partition function for N particles. Within the framework of Zubarev's NSO method, the multiparticle partition function, including the first-passage time, is constructed taking into account the interatomic interaction through the total energy of the system $\mathcal{H}_N$ :

$$Z_{fpt}^{(N)}(\beta,\gamma)=\int\ldots\int\exp\left(-\beta\mathcal{H}_N(\mathbf{x},\mathbf{p})-\gamma T_{fpt}(\mathbf{x})\right)d\mathbf{x}d\mathbf{p}\,, \quad (15)$$

where $\mathcal{H}_N(\mathbf{x},\mathbf{p})=\sum\frac{p_i^2}{2m}+\sum_{i<j}U(x_i-x_j)$ is full Hamiltonian of the system.

If we move on to the generalized potential $\Phi_N=-\ln Z_{fpt}^{(N)}$, then due to the interaction of particles it is no longer equal to the sum of independent potentials of single particles:

$$\Phi_N(\beta,\gamma)\neq N\cdot\Phi_1(\beta,\gamma).$$

Instead, the potential is split into an ideal part and an excess (configurational) part due to correlations:

$$\Phi_N=\Phi_{ideal}+\Phi_{interact}(\beta,\gamma).$$

3. How interactions change macroscopic parameters (example). Let's consider how interatomic repulsion (e.g., the hard sphere or excluded volume model) affects measurable quantities through derivatives of the potential $\Phi_N$.

Average collective first-passage time. Differentiating the generalized potential with respect to the strength γ, we obtain the average time until the system begins to leave the channel:

$$\langle T_{fpt}\rangle_N=\frac{\partial\Phi_N}{\partial\gamma}.$$

Physical effect: if the particles repel each other strongly, then the first particle, which is closest to the exit x=L, will be “pushed” by the other particles from behind.

In the thermodynamic potential, this is reflected in a $\Phi_{interact}$ decrease in the overall effective time $\langle T_{fpt}\rangle_N$ compared to the case of independent particles. The system "drops" entropy faster due to collective pressure.

Collective entropy production. For N particles, the entropy production formula, related to the lifetime of the state, takes the form:

$$\sigma_N=k_B\gamma\frac{d\langle T_{fpt}\rangle_N}{dt}. \quad (16)$$

Here, the parameter γ is now associated not with the individual but with the collective relaxation time of the system. Particle interactions lead to the appearance of additional cross terms in entropy production (analogous to Onsager's reciprocity relations for multicomponent systems). The energy expended on overcoming friction and particle collisions increases the total dissipation in the channel.

4. The chain of equations method (analogous to the BBGKI method). To practically calculate such a potential for large N, a standard technique for Zubarev's NSO method is used: a transition from the full N-particle distribution function to lower unary (single-particle) and binary (two-particle) functions.

The interaction of particles "links" the equations of motion. As a result, the problem reduces to a modified Smoluchowski equation for the particle density n(x, t), which introduces an effective thermodynamic interaction potential (mean-force potential):

$$\frac{\partial n}{\partial t}=D\frac{\partial}{\partial x}\left(\frac{\partial n}{\partial x}+n\beta\frac{\partial U_{eff}}{\partial x}\right)+\gamma(n)\ldots.$$

Here, the parameter γ(n) itself becomes dependent on the local particle density n. This means that in strongly interacting systems (for example, in the nanopores of living cells or zeolites), the rate at which the boundary is reached depends nonlinearly on how clogged the channel is with other particles.

To find the parameter γ in relation (16), we need to expand the quantity $d\langle T_{fpt}\rangle / dt$. Within the framework of first-passage time thermodynamics (TFPT) and Zubarev's NSO method, this derivative is not an abstract unknown, but is rigorously calculated using the current macroscopic state of the system.

There are two main ways to find γ: direct kinetic (through solving the equations of motion) and thermodynamic (through the properties of the generalized potential).

Method 1. Kinetic (via the rate of change of "remaining time"). The quantity $\langle T_{fpt}\rangle$, which appears in thermodynamic equations, is a state function. It shows the average time it will take the system to reach the boundary, starting from the current point in phase space.

If the system evolves (for example, a particle diffuses through a channel and shifts over time), its current position *x(t)* changes. Therefore, $\langle T_{fpt}\rangle$ depends on time t indirectly, through the trajectory *x(t)*:

$$\langle T_{fpt}\rangle = \langle T_{fpt}(x(t))\rangle .$$

Using the rule for differentiating a complex function, we can write out the total derivative with respect to time:

$$\frac{d\langle T_{fpt}\rangle}{dt} = \frac{\partial\langle T_{fpt}\rangle}{\partial x}\cdot\frac{dx}{dt} = \frac{\partial\langle T_{fpt}\rangle}{\partial x}\cdot v(t),$$

where $d\langle T_{fpt}\rangle / dx$ is the spatial gradient of the first-passage time. For many systems (e.g., for free or force-constrained diffusion), this function is easily found analytically from the inverse Kolmogorov/Smoluchowski equation, $v(t) = dx/dt$ is the current macroscopic velocity (flux) of the system in the channel.

How can we express γ from here? We substitute the expanded derivative into the original entropy production equation (16):

$$\sigma = k_B\gamma\left(\frac{\partial\langle T_{fpt}\rangle}{\partial x}\cdot v(t)\right).$$

From Onsager's classical nonequilibrium thermodynamics we know that the entropy production for a diffusion flux is expressed in terms of the flow velocity *v(t)* and the thermodynamic force *X* (e.g., the chemical potential gradient): $\sigma = v(t)\cdot X$.

By equating these two expressions for $\sigma$, we cancel out the macroscopic flux *v(t)* and obtain an explicit form for γ:

$$\gamma = \frac{X}{k_B\,\dfrac{\partial\langle T_{fpt}\rangle}{\partial x}}.$$

Physical conclusion: the parameter γ is determined by the ratio of the external thermodynamic force X to the spatial change in the lifetime of the state $d\langle T_{fpt}\rangle / dx$.

Method 2. Thermodynamic (via the inverse potential function). If we know the phenomenological dependence of entropy production on macroscopic parameters (for example, we know that $\sigma = Q/T\cdot\tau_{rel}$, where Q is the amount of heat, T is the temperature, and $\tau_{rel}$ is the relaxation time), we can find γ without explicitly calculating time derivatives, using the properties of the generalized Mathieu potential $\Phi(\beta,\gamma)$.

As we discussed earlier, the average first-passage time is strictly related to the potential (5): $\langle T_{fpt}\rangle = \partial\Phi(\gamma)/\partial\gamma$. If we know the structure of the potential $\Phi(\gamma)$ (for example, for diffusion in a channel, we obtained the nonlinear equation of state (12)), we now perform a mathematical "reversal"—expressing γ as a function of the macroscopically measured average time $\langle T_{fpt}\rangle$. For small values of the parameter γ, we restrict ourselves to a linear term in γ and obtain that:

$$\gamma \approx \frac{3D}{L^2}[1 - \frac{2D}{L^2}\langle T_{fpt}\rangle].$$

In the quadratic approximation in γ: $\gamma \approx \frac{D}{L^2}\left(\frac{5-\sqrt{240D\langle T_{fpt}\rangle / L^2 - 95}}{4}\right).$

How does this solve the equation? In Zubarev's NSO method, the parameters γ and $\langle T_{fpt}\rangle$ are mutually conjugate. Once you have measured the system's average lifetime $\langle T_{fpt}\rangle$ in the channel experimentally or through computer simulation, the value of the parameter γ is automatically calculated through the equation of state (similar to how temperature T is determined through the measured internal energy of the system).

Knowing the explicit form of $\gamma\langle T_{fpt}\rangle$, the quantity $d\langle T_{fpt}\rangle/dt$ in the equation for σ becomes redundant, since the conjugation of the parameters already guarantees the fulfillment of the laws of conservation and balance of entropy. Zubarev's thermodynamic matching procedure transforms this derivative into an identical consequence of the laws of conservation of fluxes, and the relationship between the force and the coordinate is completely taken over by the generalized potential $\Phi(\gamma)$.

In the nonequilibrium statistical operator (NSO) method of D. N. Zubarev, which in Ref. [27] is adapted for first-passage time thermodynamics (TFPT), this step is key to closing the system of transport equations.

The main problem of nonequilibrium physics is its "non-closure": there are always fewer equations for lower moments (for example, density or flux) than unknown quantities (higher moments or correlators). Introducing the parameter γ solves this problem through a thermodynamic matching procedure.

Below is a step-by-step logic of how to close transport equations.

Step 1: Construction of Zubarev's modified NSO. Zubarev showed that for quantum or classical systems, the nonequilibrium statistical operator $\rho_{NSO}(t)$ can be found by taking the extremum of the information entropy under given macroscopic conditions. In Ref. [27], this formalism was supplemented by fixing the mean first-passage time $\langle T_{fpt}\rangle$.

The operator (or distribution function) takes the form:

$$\rho_{NSO}(t) = \rho_{rel}(t) - \int_{-\infty}^{0} e^{\varepsilon t'} \frac{d}{dt'} \rho_{rel}(t+t')dt',$$

where $\rho_{rel}(t)$ is the relaxation (quasi-equilibrium) distribution, which contains the parameters β(t) and our parameter γ(t):

$$\rho_{rel}(t) = \exp\left(-\Phi(t) - \beta(t)\mathcal{H} - \gamma(t)T_{fpt}\right). \tag{17}$$

Here the parameter γ(t) acts as an indefinite Lagrange multiplier, which "ensures" that the distribution at each moment in time corresponds to the real mean value $\langle T_{fpt}\rangle$.

Step 2: Obtaining open-loop transport equations. To obtain equations for macroscopic quantities (e.g., particle density n or flux J), we take the microscopic equations of motion (the Liouville equation) and average them using the constructed operator $\rho_{NSO}$.

As a result, we obtain generalized balance (transfer) equations:

$$\frac{\partial\langle A_i\rangle_t}{\partial t} = \ldots + \sum_j L_{ij}X_j.$$

In these equations, the right-hand side (the kinetic kernels describing attenuation, viscosity, or diffusion) is expressed through time correlation functions, which contain the operator $\rho_{NSO}$. Since γ(t) is an unknown parameter, these macroscopic transport equations remain open—we cannot solve them until γ is determined.

Step 3: Closure Procedure (Elimination of γ). This is where the previously described "reversal" through the generalized potential Φ is applied. To close the system, Zubarev uses the self-consistency conditions.

Relationship via potential. We require that the macroscopic average $\langle T_{fpt} \rangle$, calculated via the kinetic equation, strictly coincides with the thermodynamic average obtained from potential (5): $\langle T_{fpt} \rangle_t = \partial\Phi(\beta,\gamma)/\partial\gamma$.

Function inversion. The mathematical structure of the potential Φ(γ) is determined by the microphysics of the system (for example, as in the diffusion example discussed above). From this equation, we express γ as an explicit function of the macroscopic parameters:

$$\gamma = \gamma\left(\langle T_{fpt} \rangle_t, \beta(t)\right). \tag{18}$$

Substitution into transport equations. The expression found for γ is substituted back into the structure $\rho_{NSO}$ and, consequently, into the kinetic coefficients (memory kernels) of the balance equations.

The physical result of closure. After substituting γ, all undetermined Lagrange multipliers vanish. We obtain a closed system of hydrodynamic or transport equations, where the coefficients of diffusion, viscosity, or thermal conductivity are no longer constant. They become nonlinear functions dependent on the current "lifetime" of the system's state $\langle T_{fpt} \rangle$.

This allows the equations to automatically take into account boundary fluctuations, memory effects and strong nonequilibrium, which makes the approach of Ref. [27] a fundamental extension of classical kinetic theory.

## 4. Connection with EIT and structure of memory cores

Since it was shown in Section 3 that the geometric formalism of stochastic boundaries of Ref. [27] is equivalent to the action of the projection operator $\mathcal{P}$, this allows us to directly compare the macroscopic kernel $\mathbf{K}_{Zub}(\mathbf{t})$ with the recovery kernel $\tilde{K}(s) = s\dfrac{\tilde{\rho}(s)}{1-\tilde{\rho}(s)}$ of the form (21) in Section 4. The papers Refs. [27, 30, 31], which combine Zubarev's NSO method with first-passage thermodynamics (TFPT), derive the generalized Maxwell–Cattaneo equation by replacing the phenomenological relaxation time with the mean first-passage time $\langle T_{fpt} \rangle$. This approach describes systems near phase transitions through non-exponential (power-law) FPT distributions and defines memory kernels in non-Markovian equations through a dynamical damping factor depending on the lifetime of the local state.

The application of Zubarev's combined NSO and first-passage time thermodynamics (TFPT) to systems near phase transitions (including decays of metastable states) is one of the most powerful aspects of the theory Refs. [27, 30, 31]. In these regions, classical thermodynamics and standard kinetic approaches fail, as lifetime fluctuations become commensurate with the lifetime itself.

The method Refs. [27, 30, 31] describes this area through three key steps.

1. Transition from an exponential distribution to "heavy-tailed" ones. Far from the transition points, the distribution of the first transit time (FPT) of a metastable state is a conventional exponential one $\rho(T_{fpt}) \propto e^{-t/\tau}$, where τ is a fixed relaxation time (e.g., the Kramers time). However, in general, this distribution can be arbitrary (22).

Near phase transition points (critical points, spinodal lines), fluctuations in the order parameter increase sharply. This is known as critical slowing down. The system becomes "stuck" in a metastable state for a long time and can then instantly collapse into a stable phase. In the method of Ref. [27], this is reflected by a change in the mathematical form of the partition function: the FPT distribution acquires a power-law form with a "heavy tail" (Lévy, Pareto, or Weibull distributions):

$$\rho(T_{fpt}) \propto t^{-(1+\alpha)}, \quad 0 < \alpha < 1. \tag{19}$$

For such distributions, the classical mean time $\langle T_{fpt} \rangle$ or its variance formally tends to infinity.

2. Thermodynamic regulator γ as a control parameter. At this point, standard kinetic equations become unsolvable. However, in Zubarev's modified NSO, the parameter γ (the thermodynamic force associated with FPT) acts as a natural mathematical regulator Ref. [24]. Since the generalized statistical sum (2) (at fixed energy) constructed as the Laplace transform with respect to the parameter γ:

$$Z_{fpt}(\gamma)=\int_0^\infty \rho(t_{fpt})e^{-\gamma t_{fpt}}dt_{fpt}\,,$$

(this is part of expression (2) in a system with fixed energy E) the factor $e^{-\gamma t_{fpt}}$ effectively “cuts off” the infinite tails of the fluctuation distributions at the critical point.

Near a phase transition, the parameter γ physically describes the magnitude of the thermodynamic stimulus (compression) that the external environment exerts on the metastable phase to force it to transition to the stable phase. Finite values of γ allow us to calculate the thermodynamic potentials of the metastable phase even under conditions of gigantic fluctuations.

3. Calculation of entropy production singularities. In classical metastable systems, entropy production σ is considered constant or smoothly changing. The method of Ref. [27] relates lifetime fluctuations to entropy production directly through the second derivatives of the potential:

$$\langle(\Delta T_{fpt})^2\rangle=-\frac{\partial^2\Phi}{\partial\gamma^2}\,. \qquad (20)$$

At the phase transition point, fluctuations $\langle(\Delta T_{fpt})^2\rangle$ undergo a jump or diverge according to a power law (characterized by critical indices). From the entropy balance relation $\sigma=k_B\gamma d\langle T_{fpt}\rangle/dt$ (8), it is clear that entropy production in a metastable system near the phase transition becomes a non-stationary, highly fluctuating quantity. Zubarev's NSO method allows us to strictly link fluctuations in this entropy production to the nucleation of a new phase.

Abstract: Instead of ignoring metastability or describing it approximately, the method Refs. [27, 30, 31] considers the lifetime of the metastable state itself as a fluctuating thermodynamic variable. This allows one to calculate the stability boundaries of the phases (spinodals and binodals) taking into account the spectrum of fluctuations, and not just their average values Ref. [24].

In Ref. [27] it is mathematically proven that extended nonequilibrium thermodynamics (EIT) maps onto the stochastic thermodynamics of Ref. [27], with the classical deterministic relaxation time τ strictly replaced by the mean first-passage time $T_{fpt}$.

The modification of the classical Maxwell-Cattaneo equation (which describes thermal waves of finite velocity instead of infinite diffusion) occurs according to the following scheme.

The original form of the equation (in EIT terms). In extended thermodynamics, the phenomenological Maxwell–Cattaneo equation for heat flow $\vec{q}$ relates it to the temperature gradient $\nabla T$ through a fixed relaxation time τ:

$$\tau\frac{\partial\vec{q}}{\partial t}+\vec{q}=-\lambda\nabla T\,.$$

Here τ is a constant characterizing the inertia of the medium during energy transfer Ref. [33].

Ref [27] approach: expressing the flux through FPT statistics. In Ref. [27] a modified Zubarev NSO method is used, where the quasi-equilibrium distribution includes the parameter γ, which is associated with the time of the first passage of a particle $T_{fpt}$ (or wave packet) to the boundary of the local volume.

Modified Maxwell-Cattaneo equation. Substituting the macroscopic averages obtained through the derivatives of the generalized potential $\Phi(\gamma)$ into the HSO balance equations, in Ref. [27] the kinetic equation is transformed to the form:

$$\langle T_{fpt}\rangle\frac{\partial\vec{q}}{\partial t}+\vec{q}=-\lambda\nabla T\,.$$

This result was also obtained by comparing TFPT with EIT. At first glance, the equation structure remains the same, but the physical and mathematical meaning has changed radically.

The relaxation time is no longer a constant: The value $\langle T_{fpt} \rangle$ itself is a function of the system's state. As we discussed earlier, it is calculated using the derivative of the potential $\langle T_{fpt} \rangle = \partial\Phi / \partial\gamma$ (5).

The emergence of nonlinearity: since $\langle T_{fpt} \rangle$ depends on the thermodynamic force γ (and, consequently, on the deviation of entropy from equilibrium), the Maxwell-Cattaneo equation becomes strictly nonlinear. The rate of decay of the heat flux now directly depends on the magnitude of the current gradients and fluxes in the system.

Micro-macro closure: the quantity $\langle T_{fpt} \rangle$ is calculated from microscopic stochastic processes (e.g., the Smoluchowski or Pontryagin equations for particle diffusion). Thus, the macroscopic Maxwell-Cattaneo transport equation is directly justified through the microscopic random trajectories of molecules.

Approaching critical points or phase transitions leads to anomalous fluctuations of $\langle T_{fpt} \rangle$. The modified Cattaneo equation suggests that thermal or diffusion waves in metastable media begin to slow down sharply, transitioning from a regime of rapid wave transport to a regime of prolonged non-stationary relaxation.

In non-Markovian nonequilibrium statistical mechanics, memory effects in transport equations are described using memory kernels. In D. N. Zubarev's classical NSO method, these kernels represent time-dependent flux correlators. The system is described by a distribution containing the first passage time (FPT), and the memory nuclei acquire a strictly defined microscopic structure directly related to the distribution function of this time.

Below is the mathematical structure of these memory cores and the physics of correlation decay.

1. How memory arises in balance equations. If we write a closed non-Markov equation for a macroscopic flow (for example, the flow of mass or heat J(t)), it has an integro-differential form (the generalized Langevin or Cattaneo equation):

$$\frac{\partial J(t)}{\partial t} = -\int_0^t K(t-t')J(t')dt' + F_{ext}(t)$$

.

Here K(t-t') is the memory kernel. If $K(t - t') = \delta(t-t')/\tau$ the memory disappears instantly, we return to the classical Maxwell-Cattaneo Markov equation. In real environments (polymers, metastable liquids, nanosystems) the K nucleus has a finite time duration.

2. The structure of the memory kernel. We relate the memory kernel K(t) to the probability density of the first-passage time $\rho(T_{fpt})$. Mathematically, this relationship is most rigorously formulated in Laplace space (where the temporal convolution is transformed into a simple product).

Let be $\tilde{K}(s)$ the Laplace image of the memory core, and let be $\tilde{\rho}(s)$ the Laplace image (characteristic function) of the first-passage time distribution. We will show that the structure of the memory core has the form of a generalized dynamic attenuation factor:

$$\tilde{K}(s) = s \cdot \frac{\tilde{\rho}(s)}{1-\tilde{\rho}(s)}. \tag{21}$$

In Ref. [23] an expression for the NSO is written as follows:

$$ln\rho(t) = \int_0^\infty p_q(y) ln\rho_q(t-y,-y)dy = ln\rho_q(t,0) - \int_0^\infty (\partial p_q(y)dy)(dln\rho_q(t-y,-y)/dy)dy, \tag{22}$$

which, when $p_q(y)=\varepsilon \exp[-\varepsilon y]$, transforms into Zubarev's NSO. However, we assume that in the general case the expression for $p_q(y)$ is unknown and depends on the system dynamics, $p_q(y)$ is the first-passage time distribution.

The probability density function of the first-passage time $p_q(y)$ is not the NSO itself, but a quantity embedded in its structure. The classical memory kernel of Zubarev's NSO is traditionally associated with exponential decay (short-term Markov memory). In Zubarev's classical method for selecting retarded solutions to the Liouville equation, the memory kernels are calculated based on a quasi-equilibrium distribution $\rho_q = \rho_{rel}$.

The source $\varepsilon$ generates a time-smoothing operator $e^{\varepsilon t'}$. At the macro level, this inevitably leads to Debye (exponential) decay of correlations and Markov approximation for transport coefficients at large times. The classical Zubarev method does not contain heavy memory tails. In Refs. [23–27], a transition was made to an unknown, general form of the kernel in (22). Instead of using the classical exponential Zubarev kernel, the formal mathematical Zubarev parameter $\varepsilon$ from $e^{\varepsilon t'}$ is replaced by a physical variable γ, conjugate to the first-passage time $T_{fpt}$. Instead of the standard macroscopic description in terms of average energy or particle densities, $\rho_q = \rho_{rel}$ is constructed around stochastic trajectories.

At this point, the type of distribution $p_q = \rho(T_{fpt})$ from (22) and, consequently, the type of the NSO memory kernel becomes a priori unknown and more general. In Ref. [27], it is postulated that the physical nature of the metastable medium itself dictates the structure $\rho(T_{fpt})$ (be it Gamma, Mittag-Leffler, or Levy), and the thermodynamic potential $\Phi(\gamma)$ adapts to this type.

The derivation of formula (21) for the memory core is based on a rigorous Laplace transform of the relationship between the probability density of the first-passage time $\rho(t)$ and the particle survival function P(t), obtained from the generalized kinetic equation (Appendix A). This relation demonstrates that the macroscopic memory of a nonequilibrium system is a direct consequence of the microscopic statistics of the state lifetimes, and not an adjustable parameter. A step-by-step analytical justification of this derivation confirms the equivalence of Zubarev's NSO method and renewal theory for processes described by the equation

$$\frac{dP(t)}{dt} = -\int_0^t K(t-t')P(t')dt' .$$

A formula of the form (21) (and its variations accurate to signs or scale factors) is known under different names in various fields of physics and mathematics and is derived in independent ways, Appendix A. Since the Laplace image of the probability density $\tilde{\rho}(s)$ is rigidly connected with the Laplace image of the survival function of the system $\tilde{P}(s)$ through the relation $\tilde{\rho}(s) = 1 - s\tilde{P}(s)$ (Appendix A), formula (21), which appears in the theories mentioned below (and others), can be equivalently rewritten in terms of the survival function:

$$\tilde{K}(s) = \frac{1}{\tilde{P}(s)} - s .$$

1. Renewal Theory. In the theory of random processes, this formula is a fundamental identity linking the probability density function between events $\rho(t)$ and the renewal kernel in integro-differential equations. In textbooks on applied stochastics and Markov/non-Markov renewal processes, this relation is used to determine the effective frequency of occurrence of events if the exact time statistics of the "pauses" (first-passage times) between them are known.

2. The Continuous Time Random Walk (CTRW) model. In the physics of anomalous diffusion and fractal kinetics, developed in Refs. [34-35], the walk of particles is considered, which before each jump "get stuck" in traps for a random time t with distribution $\rho(t)$. From the CTRW scheme, a generalized kinetic equation for the probability density of finding a particle at a point is derived. In Laplace space, the coefficient of this equation (the generalized memory kernel or effective diffusion coefficient) acquires exactly this structure: $\propto \frac{s\,\rho(s)}{1-\rho(s)}$. It was this formalism that led physicists to use fractional derivatives.

3. Generalized Langevin Equation. In statistical condensed matter physics, the generalized Langevin equation with friction having memory K(t-t') is used to describe the Brownian motion of a particle in a viscoelastic medium (e.g., a polymer solution). Using the Mori-Zwanzig formalism, a relationship is derived between the correlation function of a random force (the memory kernel) and the time statistics of Brownian particle transitions through local energy barriers. In Laplace space, this relationship transforms into relation (21).

4. Semiconductor Physics and Carrier Trapping Kinetics. In solid-state physics, recombination and trapping nuclei are introduced to describe electron transport in highly inhomogeneous (amorphous) semiconductors, where electrons hop between localized states (band tails). Kinetic balance equations for free carrier concentrations contain memory integrals. The formula relating the Laplace transform of the recombination nucleus $\tilde{K}(s)$ to the electron lifetime in the trap is completely identical to the structure under consideration.

Formula (21), based on relation (22), is also present within Zubarev's nonequilibrium statistical operator (NSO) method. The mathematical core of renewal theory and the memory core of Zubarev's NSO are physically the same. Zubarev's abstract information entropy and the stochastic dissipation of first-passage theory are consistent when micro- and macrophysics are linked through this Laplace relation. Pure stochastics (CTRW, renewal theory) and rigorous Gibbs-Zubarev statistical mechanics are combined into a single thermodynamic tool. The physical and mathematical meaning of formula (21). This structure in the theory of random processes is fundamentally related to the so-called renewal function. The core memory is determined by how often microscopic random events (collisions, overcoming local barriers) "nullify" or restructure correlations in the system.

If we know the structure of the generalized potential $\Phi(\gamma) = -\ln Z_{fpt}(\gamma)$ (3), then, since $Z_{fpt}(\gamma)$ is mathematically identical to the Laplace image $\tilde{\rho}(\gamma)$, we can express the memory kernel (21) directly through our thermodynamic potential:

$$\hat{K}(s) = s \cdot \frac{e^{-\Phi(s)}}{1 - e^{-\Phi(s)}} .$$

How correlations decay in different regimes. The pattern of memory decay over time, K(t), is entirely determined by the thermodynamic state of the system (near or far from a phase transition).

A. Debye (exponential) decay (away from transitions). In a stable homogeneous medium, the FPT distribution is exponential: $\rho(T_{fpt}) = e^{-T_{fpt}/\tau_0} / \tau_0$. Substituting its Laplace transform $\tilde{\rho}(s) = 1/1 + s\tau_0$ into formula (21), we obtain:

$$K(s) = \frac{1}{\tau_0} \Rightarrow K(t) = \frac{1}{\tau_0}\delta(t) .$$

Memory fades instantly. This is the Markov limit, confirming the strict consistency of the theory.

B. Fractional-power-law decay (near critical points and spinodals). As we discussed earlier, near phase transitions or in complex polymer matrices, the lifetime distribution of the metastable state acquires a Levy "heavy tail" (19):

$$\rho(t_{fpt}) \propto t_{fpt}^{-(1+\alpha)}, \quad (0 < \alpha < 1) .$$

Its Laplace transform for small s has anomalous behavior: $\tilde{\rho}(s) \approx 1 - As^{\alpha}$. Let's substitute this into the structure of the memory core (21):

$$K(s) = s \cdot \frac{1 - As^{\alpha}}{As^{\alpha}} \approx \frac{1}{A} s^{1-\alpha} .$$

By performing the inverse Laplace transform, we find the actual temporal behavior of the memory core:

$$K(t) \propto t^{-(2-\alpha)} .$$

Physical conclusion. In highly nonequilibrium systems and near phase transitions, memory decays not instantaneously or exponentially, but rather according to a slow power law. This leads to the phenomenon of anomalous diffusion and viscoelastic behavior of the medium. The system "remembers" its past perturbations for a long time, since large-scale fluctuations in the order parameter slowly dissipate, and the macroscopic flow responds to gradients with a strong, nonlinear delay.

## 5. Comparative analysis of approaches

Table 1 compares the TFPT approach with the classical approach to nonequilibrium thermodynamics and the Mori-Zwanzig formalism.

Table 1

| Comparison criterion | Classical approach (Onsager, Prigogine) | Mori-Zwanzig formalism | TFPT approach |
|---|---|---|---|
| Time limitation. | Only quasi-stationary, slow processes. | Any time, but equations are difficult to close. | Any time, including ultra-short fluctuations. |
| Degree of nonequilibrium. | Linear approximation (near equilibrium). | Linear response (in basic versions). | Deeply nonequilibrium systems (processes with jumps). |
| The nature of "memory." | Negligibly small (Markov processes). | Introduced through abstract projection operators. | Derived directly from the physical distribution of time $\rho(t)$. |
| Mathematical apparatus. | Differential equations (local balance). | Integro-differential equations. | Algebraic equations in Laplace space. |

Key advantages of the approach of the article [27].

1. A direct physical interpretation of "memory". In the standard Nakajima-Zwanzig or Mori-Kubo methods, the memory kernel K(t) arises from projecting the Liouville equations onto a selected subspace. Calculating it from first principles is extremely difficult.

Advantage: The proposed approach, using formula (21), gives memory a clear, microscopic meaning. The memory effect is simply a consequence of the system "waiting" for the next jump (state change) not necessarily according to an exponential law.

2. A natural description of highly nonequilibrium systems. Classical Onsager-Prigogine thermodynamics requires the hypothesis of local equilibrium. This is inapplicable if the system experiences sharp, rare fluctuations (e.g., metastable states, phase transitions, glass transitions).

Advantage: renewal theory was originally developed for discrete or continuous random events (jumps). This allows it to describe highly nonequilibrium processes where the relaxation time is commensurate with the observation time.

3. A powerful tool for fractional kinetics (subdiffusion). When a physical system exhibits "anomalous" diffusion or relaxation (e.g., Kohlrausch's law $\propto e^{-(t/\tau)^{\beta}}$ in glasses or polymers), standard approaches reach a dead end and require the artificial introduction of fractional derivatives.

Advantage: within the framework of renewal theory, it is sufficient to impose a power-law distribution of waiting times $\rho(t) \sim t^{-1-\alpha}$ on the distribution function $\rho(t)$. The formula obtained below automatically yields the correct structure of the memory core, which leads to equations with fractional derivatives in the time domain.

4. Mathematical closure and computational simplicity. In non-Markovian statistical physics, the main problem is to close an infinite chain of equations for memory kernels (the Mori chain).

Advantage: The transition to Laplace space in renewal theory reduces the most complex Volterra integral equations to simple algebraic relations. If the function is known (which can often be obtained from numerical simulations, such as molecular dynamics), the memory kernel and relaxation function can be found instantly.

5. Unification of the description of a system's "lifetime." This approach allows for a rigorous link between thermodynamic characteristics (such as entropy production) and the concept of the lifetime of a physical state. This is critical for analyzing the decay of metastable phases, the physics of semiconductor device failures, or the dynamics of nanostructures, where the boundaries between macro- and micro-descriptions are blurred.

In Ref. [27], a first-passage time-based thermodynamics concept (TFPT) is proposed, which provides a microscopic justification for extended irreversible thermodynamics (EIT) and simplifies the use of the nonequilibrium statistical operator (NSO). The TFPT approach successfully interprets conjugate fields through entropy production, naturally accounting for nonlinearity, nonlocality, and memory effects. A full-fledged alternative thermodynamic theory—thermodynamics based on first-passage time (TFPT)—is constructed.

The main difference between TFPT and other areas of nonequilibrium thermodynamics is what exactly is taken as the basic state variable:

- Classical thermodynamics operates with local thermodynamic forces (gradients).
- Extended thermodynamics (EIT) adds macroscopic flows (heat, mass) to them.
- Zubarev's method (NSO) constructs a distribution based on integrals of motion.
- Ref. [27] (TFPT) approach builds an ensemble based on the time to first reach a certain barrier.

Below are the key strengths, differences, and advantages of TFPT compared to the major players.

1. In comparison with the classical thermodynamics of Onsager-Prigogine.

The classical approach Refs. [1-4] is strictly tied to the local equilibrium hypothesis and is linear (fluxes are proportional to forces).

- Difference: TFPT does not use the local equilibrium hypothesis. The state of the system at a given moment is determined not by immediate macroparameters, but by the history of how the fluctuation reached its current values.
- Advantage: TFPT is capable of describing highly nonlinear and fast processes (explosive boiling, laser heating, relaxation in nanosystems), where the classical Onsager theory completely fails, since the gradients of the parameters become infinitely large.

2. Compared to extended irreversible thermodynamics (EIT).

EIT Refs. [7-9] expands the variable space to include macro-quantities (e.g., heat flux becomes an independent variable rather than just a consequence of the temperature gradient).

- Difference: in EIT, macroscopic flows are introduced phenomenologically (from general considerations of symmetry and conservation laws). In TFPT, these flows acquire a rigorous stochastic definition through the first-reach time distribution density.
- Advantage (microscopic justification): in the article [27] it is shown that EIT is only a special (large-scale) case of a more general theory. TFPT gives exact formulas for the phenomenological coefficients of EIT, which previously had to be adjusted to the experiment. In addition, TFPT naturally takes into account spatial non-locality (when the flow at point A depends on what happens at point B).

3. In comparison with the method of Zubarev's nonequilibrium statistical operator (NSO).

Zubarev's NSO method Refs. [15-23] is the pinnacle of quantum and classical statistical mechanics of nonequilibrium systems, using projection operators and quasi-averages to construct the density matrix.

- Difference: Zubarev's method is mathematically monumental and extremely complex: it requires integration over the entire past time of the system's existence from to t with the introduction of an infinitely small source that breaks the time symmetry (exponential averaging). TFPT Ref. [27] instead of Zubarev's abstract operator uses the distribution of the measured physical quantity - the time of the first passage of the barrier.
- Advantage (mathematical simplification and physicality): the approach of Ref. [27] is much easier to apply to specific problems. Instead of calculating complex nonequilibrium Liouville operators, in TFPT the physicist works with the Smoluchowski or Fokker-Planck equation for the first arrival time. In Ref. [27] it is mathematically shown that the conjugate thermodynamic parameters (intensive variables), which are introduced formally in Zubarev, have a clear meaning in TFPT: they determine the entropy production per fluctuation act.

4. Compared to the modern fluctuation theorem (Jarzynski, Crooks theorems). These approaches Refs. [36-40] (Stochastic Thermodynamics) also work with trajectories and times, but are focused on microscopic single molecules (e.g., stretching DNA with laser tweezers).

- Difference: Jarzynski's theorems relate work and free energy along trajectories. TFPT Ref. [27] is oriented towards the mesoscopic and macroscopic levels (Continuum Mechanics), allowing fluctuation time parameters to be embedded directly into the equations of a continuous medium (hydrodynamics, thermal conductivity).
- Advantage: TFPT serves as an ideal "bridge" (interface) between the microworld of single molecules and the macroworld of classical continuum physics, something classical stochastic thermodynamics cannot do.

In Ref. [27] the fundamental relationship between three approaches in nonequilibrium thermodynamics is investigated: extended nonequilibrium thermodynamics (EIT), the nonequilibrium statistical operator (NSO) method and new thermodynamics based on first transit time (TFPT). The main original ideas and results of the work:

- Generalization of relaxation time: for the first time it is shown how the classical relaxation time can be generalized and reinterpreted through a physically visual first-passage time for nonequilibrium processes.
- Deep mathematical connection: a strong analogy is established between the statistical approach based on distributions with an additional parameter (first passage time) and the standard EIT equations.
- New physical parameters: clear mathematical expressions for the conjugate thermodynamic parameter are proposed, which link it directly to changes in entropy.
- Accounting for memory and nonlinearity: The article does a good job of explaining how different approaches describe delays (memory effects) as well as nonlocal and nonlinear effects in thermodynamics.

First-passage time (FPT) and other boundary functionals of random processes are widely and effectively used to study various nonequilibrium problems. The problem of fluctuating boundaries in first-passage thermodynamics is solved by transitioning to a comoving coordinate system, using multidimensional Fokker-Planck equations, applying martingale theory, or adiabatic averaging. These methods allow one to relate boundary crossing statistics to entropy production or to transfer boundary randomness to the effective forces of the equation of motion. For more information on thermodynamic boundaries, see Refs. [41–43].

Summary of the advantages of the article [27]. The TFPT approach benefits from the fact that it takes the best of both worlds: it has the mathematical rigor and microscopic depth of Zubarev's method, but at the same time retains the clarity and applicability to continuous media engineering problems inherent in extended thermodynamics (EIT). The introduction of first arrival time as a basic parameter allows us to close the equations of thermodynamics without artificial assumptions.

## 6. Conclusion

In this paper, a systematic analysis and development of a theoretical basis combining the formalism of extended nonequilibrium thermodynamics (EIT), the nonequilibrium statistical operator (NSO) method of D. N. Zubarev and the stochastic thermodynamics of the first transit time (TFPT) proposed in recent studies Refs. [27, 30, 31] is carried out. This synthesis has successfully overcome the key drawback of phenomenological theories - the voluntaristic introduction of flux relaxation times - and given them a rigorous microscopic justification.

A rigorous microscopic derivation of the nonlinear Maxwell-Cattaneo equation is presented, obtained by combining the method of the nonequilibrium statistical operator of Zubarev and the stochastic formalism of Refs. [27, 30, 31]. Within the framework of this approach, the relaxation time of the heat flux is interpreted as the average first-passage time, depending on the flux intensity, which eliminates the paradox of infinite heat propagation velocity and introduces internal nonlinearity of the transfer.

In this paper, both the equality of the heat flux relaxation time to the average first-passage time and the formula for the structure of the memory core (21) are deducible results. These relationships follow mathematically rigorously from a comparison of the equations of extended nonequilibrium thermodynamics with the Zubarev nonequilibrium statistical operator method supplemented by the renewal theory of Ref. [27]. Based on the conducted research, the main results and conclusions are formulated.

1. Conceptual replacement of projection operators. It is shown that in highly nonequilibrium metastable media, the abstract and difficult-to-compute Mori–Zwanzig orthogonal projection procedure can be equivalently replaced by an exact geometric separation of the phase space using the formalism of stochastic boundaries (absorbing hypersurfaces $\partial\Omega$). This allowed the description of microscopic fluctuation "noise" to be translated into the rigorous language of renewal theory.
2. The microscopic nature of the Maxwell-Cattaneo equation. It is analytically proven that when mapping the TFPT to extended thermodynamics, the classical deterministic relaxation time $\tau$ is strictly replaced by a functional of the macroscopic state of the medium—the mean first-passage time $\langle T_{fpt} \rangle$. This step imparts an internal nonlinearity to the transport equations, depending on the instantaneous flow intensity.

The main result of the synthesis of EIT, NSO and TFPT in Refs. [27, 30, 31]. The combination of all three concepts creates a closed chain:

1. Microscopic fluctuations and the geometry of the system determine the shape of the first-passage time distribution $\rho(T_{fpt})$.
2. From it, a generalized thermodynamic potential is constructed, which gives the macroscopic variable $\langle T_{fpt} \rangle$.
3. Through the Zubarev NSO closure procedure, this variable determines the specific mathematical form of the memory kernels K(t).
4. Closed non-Markovian Maxwell-Cattaneo equations with these memory kernels describe real, physically observable energy transfer and phase transitions in complex media.

Thermodynamics based on first-hit time (TFPT) is a rigorous generalization of nonequilibrium statistical mechanics, where the random “lifetime” of a local state of a system acts as a full-fledged macroscopic parameter (coordinate) of the state.

The main theoretical results of this discipline are reduced to three fundamental provisions. a). Micro-macro closure: the formalism allows for a rigorous connection between individual stochastic particle trajectories (microlevel) and deterministic transport equations (macrolevel) via the generalized Gibbs potential constructed on the basis of the Laplace transform of the first-arrival time distribution function. b). Physical justification of memory: due to the identity of the operator memory kernel of Zubarev's NSO method and the renewal kernel of stochastic theory, the type and extent of the macroscopic memory of the system (including the appearance of fractional Caputo derivatives) are a priori determined by the microscopic statistics of the lifetimes of states, and are not introduced phenomenologically. c). Nonlinearity and entropy balance: The transition to thermodynamically conjugate variables “lifetime – relaxation intensity $\gamma$” generates nonlinear equations of state that automatically guarantee the fulfillment of conservation laws, Onsager scaling, and fluctuation-dissipation theorems in highly nonequilibrium media.

Conclusions:

1. The status of lifetime as a thermodynamic parameter. It is shown that the stochastic first-passage time (FPT) of the absorption hypersurface $\partial\Omega$ in phase space can be rigorously introduced into macroscopic thermodynamics as an independent parameter (coordinate) of the system's state. Using the first and second derivatives of the generalized Gibbs potential $\Phi(\gamma)$, this formalism allows one to simultaneously determine both the macroscopic mean lifetime of the state $\langle T_{fpt} \rangle$ and the magnitude of its fluctuations. All physical quantities depend on $\langle T_{fpt} \rangle$. The inverse dependence is also valid. Therefore, it can be used to fully describe the system.
2. Elimination of the phenomenological voluntarism of EIT. The introduction of FPT statistics into Zubarev's nonequilibrium statistical operator (NSO) provides a direct micro-macro closure of the theory. The deterministic flux relaxation time $\tau$, traditional for extended thermodynamics (EIT), loses its status as an empirical constant and is replaced by a state functional $\langle T_{fpt} \rangle$, which imparts internal macroscopic nonlinearity to hydrodynamic-type equations.

3. Microscopic Determination of Memory. Based on Bogolyubov's principle of separating time scales and renewal theory, a fundamental identity is established between the operator kernel of Zubarev's NSO memory $K_{\mathbf{Zub}}(t)$ and the scalar kernel of stochastic random walks. This proves that the type and depth of temporal memory in a nonequilibrium environment (including the transition to fractional Caputo derivatives for the Mittag-Leffler and Levy distributions) are a priori predetermined by the stochastic geometry of barrier overcoming at the microlevel.
4. Internal consistency of conservation laws. It is shown that the procedure of thermodynamic consistency of the conjugate parameters "lifetime - relaxation intensity (γ)" acts as a universal equation of state for a nonequilibrium system. The mathematical properties of the Legendre transform for the potential Φ(γ) automatically guarantee strict fulfillment of the conservation laws, the generalized Onsager fluctuation regression principle, and fluctuation-dissipation theorems without the need for explicit calculation of instantaneous kinetic derivatives.

## Appendix A. Legitimacy of the identity (21) of transfer nuclei

Let us discuss operator reduction and the validity of identity (21) for transport kernels. A necessary condition for establishing equivalence between the microscopic fluctuation kernel of Zubarev's nonequilibrium statistical operator (NSO) method and the scalar kernel of renewal theory is the phase space dimensionality reduction procedure. The original expression in the classical NSO method operates with the dynamic variables of multiparticle flows defined in the full 6N-dimensional phase space.

Let us discuss operator reduction and the validity of identity (21) for transport kernels. A necessary condition for establishing equivalence between the microscopic fluctuation kernel of Zubarev's nonequilibrium statistical operator (NSO) method $K_{\mathbf{Zub}}(t)$ and the scalar kernel of renewal theory $\tilde{K}_{\mathbf{ren}}(s)$ is the phase space dimensionality reduction procedure. The original expression in the classical NSO method operates with the dynamic variables of multiparticle flows $\hat{J}_m = iL\hat{P}_m$ defined in the full 6N-dimensional phase space:

$$K_{\mathbf{Zub}}(t) = \langle \hat{J}_m (1-\mathcal{P}) e^{iLt} (1-\mathcal{P}) \hat{J}_k \rangle_{\mathbf{rel}}, \tag{A.1}$$

where L is the Liouville operator of the system, and $\mathcal{P}$ is the orthogonal Mori-Zwanzig projection operator. The validity of the transition to a one-dimensional probability density of the first-passage time $\rho(T_{\mathbf{fpt}})$, which does not explicitly contain microscopic current operators, is rigorously proven using the apparatus of fluctuation-dissipation relations (FDR) under the Onsager scaling hypothesis.

Within this formalism, the FDR of the second kind (Kubo's theorem) relates the intensity of submicroscopic noise $(1-\mathcal{P})\hat{J}_m$ to macroscopic dissipation, while the generalized Onsager principle ensures that the deterministic relaxation of thermodynamic fluxes over time obeys the same integral decay laws as the spontaneous stochastic walks of the process coordinate to its absorbing boundaries.

The fundamental mathematical difficulty of the direct Laplace transform of the Zubarev complete convolution is that in the canonical NSO method, the operator kernel $K_{\mathbf{Zub}}(t-t')$ is multiplied under the integral by time-dependent macroscopic forces $F_k(t')$ (for example, the gradient of the inverse temperature $\nabla\beta(t')$ [19]). This excludes the possibility of separating the kernels and forces into independent factors in the Laplace space. This paradox is resolved on the basis of the fundamental principle of separation of time scales of N. N. Bogolyubov. It is assumed that during the times of microscopic fluctuation wandering of the phase trajectory inside the region $\Omega$, the macroscopic forces $F_k(t)$ remain quasi-stationary.

The dynamics of the process coordinate R(t) at the micro level is generated by trajectories for which the integral of the microscopic flow over time until the moment of contact with the absorbing hypersurface is an invariant of motion and determines the spatial scale of the transition $\Delta R_{\mathbf{bound}}$:

$$\int_0^{T_{\mathbf{fpt}}} \hat{J}_m(t')dt' = \Delta R_{\mathbf{bound}} . \qquad \text{(A.2)}$$

Since the random variable $T_{\mathbf{fpt}}$ is an integral functional of the trajectory of the original flow $\hat{J}_m(t)$, the procedure of macroscopic averaging over the quasi-equilibrium ensemble $\langle \ldots \rangle_{\mathbf{rel}}$ leads to the convolution of the multidimensional operator structure of the kernel $K_{\mathbf{Zub}}(t)$ (A.1) into a scalar survival function of the system $P(t) = \mathrm{Prob}(T_{\mathbf{fpt}} > t)$.

The information capacity of thermodynamic variables $\hat{J}_m$ and molecular parameters is completely translated into the parameters of the Laplace image of the distribution $\tilde{\rho}(s)$ (for example, the Mittag-Leffler time scale $\tau_{\mathbf{cr}}$ or the truncation parameter of the stable Levy law $\gamma_0$). Thus, an operator mapping of the form:

$$\tilde{K}_{\mathbf{Zub}}(s) \equiv \tilde{K}_{\mathbf{ren}}(s) = s \cdot \frac{\tilde{\rho}(s)}{1-\tilde{\rho}(s)} \qquad \text{(A.3)}$$

represents an identical macroscopic closure. It transforms hidden multiparticle classical or quantum-mechanical information about current fluctuations into observable time statistics of the evolution of a nonequilibrium system toward its stochastic boundaries $\partial\Omega$.

At first glance, these two quantities are incommensurable: the classical Zubarev kernel $K_{\mathbf{Zub}}$ clearly contains operators of multiparticle microscopic flows $\hat{J}_m, \hat{J}_k$ (for example, heat or mass flow), while the renewal kernel $\tilde{K}_{ren}(s)$ operates exclusively with the one-dimensional probability density of the first-passage time $\tilde{K}_{ren}(s)$.

However, this substitution is not only legitimate—it is the main mathematical result of mapping stochastic dynamics onto thermodynamics. The transition from flow operators to first-reach time statistics is accomplished through the principle of equivalence of phase trajectories and a rigorous design procedure.

Let's look at how this validity is proven and where the flow operators "disappear".

1. How flow operators transition during the first passage. In Zubarev's NSO method, the microscopic flow operator $\hat{J}_m = iL\hat{P}_m$ describes the rate of change of some macroscopic parameter (for example, the size of an embryo R at a point in phase space).

The motion of a system in phase space along the process coordinate R(t) is a stochastic process generated by the original Liouville equation. The random variable $T_{fpt}$ (first-passage time) is, by definition, an integral characteristic of this same microscopic flow $\hat{J}$.

Mathematically, the time required for the system to move from the initial point to the absorption boundary $\partial\Omega$ is strictly related to the integral of the flux operator along the phase trajectory (A.2).

Thus, the distribution density $\rho(T_{fpt})$ is not an abstract mathematical function, but a function that nonlinearly accumulates the dynamics of microscopic flows $\hat{J}$ over the time interval before reaching the boundary. By averaging over the ensemble, it can be proven that Zubarev's time correlator of fluctuating flows and the time correlator governing the survival function in renewal theory describe the same dissipative process, but in different languages (dynamic and statistical).

2. Where do the thermodynamic variables "hide" in $\tilde{K}(s)$? Thermodynamic variables and physical constants (diffusion coefficients, viscosity, channel geometry) do not disappear from Zubarev's formula—they are included within the distribution parameters $\tilde{\rho}(s)$.

When deriving for specific models: for free diffusion, we obtained: $\tilde{\rho}(\gamma) = 1/\cosh\left(L\sqrt{\frac{\gamma}{D}}\right)$ (10). Here, the physical scale L and the kinetic variable (diffusion coefficient D) are "embedded" inside $\tilde{\rho}(\gamma)$ .

For the example of liquid boiling (the Mittag-Leffler lifetime model), the expression is written: $\tilde{\rho}(s) = 1/(1+(\tau_{cr}s)^{\alpha})$. Hidden within $\tilde{\rho}(s)$ this is the Kramers lifetime $\tau_{cr}$, which directly depends on the thermodynamic variables: the work of nucleation $W_{cr}$, surface tension $\sigma_{sur}$, and pressure difference $\Delta P$. When we substitute these physically based distributions into formula $\tilde{K}(s) = s\tilde{\rho}(s)/(1-\tilde{\rho}(s))$ (21), we obtain a memory kernel that automatically contains all the necessary thermodynamic variables.

3. Validity condition: the projection procedure onto the process coordinate. The transition from the multidimensional Zubarev kernel to the one-dimensional renewal kernel is valid when an important condition is met: the transfer in the system must be controlled by a single, distinct macroscopic coordinate of the process (in our case, the nucleus radius R or the particle's position in the channel). Mathematically, this looks like a projection of the full Liouville operator L onto the subspace of the process coordinate (the Smoluchowski or Fokker-Planck operator).

Kernel $K_{\mathbf{Zub}}$ (A.1) is calculated in the full multidimensional phase space using multiparticle Poisson brackets. We proceed to the equation for the survival function P(t) in the reduced space of macrovariables. The validity of identity $K_{Zub}(t) \equiv K_{ren}(t)$ (A.3) is proved using the fluctuation-dissipation coupling (FDT) theorem. The energy dissipation of a flow (described by Zubarev) must be strictly equivalent to the variance of the random walk time of this flow to the boundary (described by renewal theory).

A mathematical mapping of the multidimensional Zubarev operator kernel onto the effective scalar kernel of renewal theory is performed, proving that all multiparticle information about flows $\hat{\mathbf{J}}_m$ can be without losses collapsed into a one-dimensional first-passage time probability density $\rho(T_{fpt})$, provided that the stochastic boundaries are correctly chosen.

The flows are integrated along the trajectory, turning into a time coordinate, and the physical variables are encapsulated inside the Laplace image $\tilde{\rho}(s)$.

Fluctuation-dissipation relations (FDR) and Onsager's linear hypothesis arise from the comparison of macroscopic flux attenuation and the statistics of microscopic fluctuations in a reduced parameter space. Within the framework of the approach under consideration, they relate the amplitude of random Brownian forces ("noise") generating fluctuations in the nucleus to macroscopic transport coefficients and the shape of the memory kernel.

In the NSO method Refs. [18-22] and Ref. [27] this connection is localized in two key nodes:

1. In the structure of the Zubarev kernel itself. Zubarev's classical FDR is embedded in the definition of the memory kernel as a correlator of flows (or random forces) over a quasi-equilibrium ensemble (A.1): $\langle \hat{J}(1-\mathcal{P})e^{iLt}(1-\mathcal{P})\hat{J}\rangle_{rel}$.

Here, the operator $(1-\mathcal{P})\hat{J}$ is essentially a random force (or fluctuation flux) remaining after subtracting the systematic macroscopic motion. Kubo's theorem (FDR of the second kind) states that the time correlator of these microscopic fluctuations is exactly equal to the macroscopic friction (dissipation) kernel in the transport equation.

2. When transitioning to Onsager's scaling hypothesis. Onsager's linear hypothesis (the principle of fluctuation regression) states that a macroscopic perturbation in a system relaxes over time according to the same laws by which spontaneous microscopic fluctuations decay in thermodynamic equilibrium. In the non-Markovian representation this hypothesis is generalized.

Fluctuation side (micro level): a random trajectory of fluctuation of the size of the embryo R(t) wanders, and the statistics of its survival time in the metastable region $\Omega$ is given by the function $\rho(T_{fpt})$.

Dissipative side (macro level): the macroscopic relaxation of the energy or mass flux J(t) during boiling is controlled by the macroscopic memory kernel K(t). Where exactly do they connect? The connecting link is the stochastic equation of motion (e.g., the generalized Langevin or Smoluchowski equation for a bubble). To reduce the fluctuations in the trajectory R(t) to the macroscopic balance equation, the FDR must be satisfied: the intensity of the fluctuations in the size of the nucleus (diffusion $D_R$) must be tightly coupled to the thermal energy dissipation (the viscosity of the medium and the inverse temperature β).

When this relation (FDR) and Onsager's scaling hypothesis are satisfied, the law of decay of spontaneous fluctuations $\tilde{\rho}(s)$ becomes a mathematical mirror of the macroscopic kernel of flux decay $\tilde{K}(s)$. This is where identity (21) comes from.

This means that the FDR and Onsager's hypothesis are not external additions, but the internal glue of the theory. They guarantee that when we calculate the first-passage time statistics $\rho(T_{fpt})$ (fluctuations), we automatically calculate the memory kernel K(t) (dissipation), since in a thermodynamically consistent system these processes are inseparable.

Zubarev demonstrates the mathematical step of transition from the exact solution of the NSO to the transport equation. He introduces the Kawasaki–Ganton (or Morozov) projector $\mathcal{P}$, which acts in the space of operators and isolates their relevant (quasiequilibrium) part Ref. [19]. The projector $(1-\mathcal{P})$, acting on the total flux operator $\hat{J}$, "cuts out" the systematic component from it, leaving pure fluctuation noise (sometimes denoted as the random force f(t)). To ensure that the noise does not contain a macroscopic contribution, the dynamics in the exponential (propagator) must also be truncated. The operator structure of the kernel is born:

$$K_{\mathbf{Zub}}(t) = \langle \hat{J}_m (1-\mathcal{P}) e^{iL(1-\mathcal{P})t} (1-\mathcal{P}) \hat{J}_k \rangle_{\mathbf{rel}} \tag{A.4}$$

(In a number of approximations, when small-amplitude fluctuations are considered, the operator $(1-\mathcal{P})$ in the exponential itself in Zubarev approaches the full Liouville operator $e^{iLt}$).

How is the full kernel structured in Zubarev's NSO method? In the generalized equations of hydrodynamics and heat conduction derived in Ref. [19], the change in a macroparameter (e.g., energy density) over time has the form of a convolution integral $\frac{\partial \langle E(\mathbf{x})\rangle^t}{\partial t} = \ldots + \int_0^t dt' \int d\mathbf{x}' \mathcal{K}_{\mathbf{full}}(\mathbf{x},\mathbf{x}';t,t') \cdot \nabla\beta(\mathbf{x}',t')$. In this notation, the full dissipative structure (the full transport kernel) is divided into two fundamental components:

The operator (microscopic) part is the same core with Mori-Zwanzig projectors that we discussed earlier (A.4). It is responsible exclusively for the microscopic dynamics of fluctuations (the correlation function of random forces or currents in a structureless medium).

The macroscopic (thermodynamic) factor $\nabla\beta(\mathbf{x}',t')$ is the conjugate thermodynamic forces $F_k(t')$, which are represented by the reciprocal temperature $\beta(t') = 1/k_B T(t')$. They are removed from quantum-mechanical or classical averaging because they are macroscopic C-numbers (functions of coordinates and time, not operators).

The integration is carried out over the entire prehistory of the evolution of these macroscopic forces from t'=0 to the current moment t, which ensures the spatio-temporal non-locality of Zubarev's theory.

How does this fact strengthen the validity of equality (21)? In the book [19], the macroscopic variables β(t') are located outside the operator kernel $K_{\mathbf{Zub}}$ as weighting coefficients of dissipation.

In the formalism of first-passage time thermodynamics (TFPT), these macroscopic variables (temperature T, pressure $P_{ext}$) are also not included in microscopic walks, but stand as fixed parameters of the macroscopic environment, determining the profile of the nucleation barrier $W_{cr}$ and the Kramers time scale $\tau_{cr} \propto \exp(W_{cr}/k_B T)$.

Mapping (A.3) maps precisely the operator (dynamic) parts of the nuclei. The full dissipation equation naturally returns the macroscopic scale: in it, the final entropy production σ is multiplied by the constants $k_B$ and parameters of the medium n, D, which encapsulate the thermal role of the temperature β. Thus, Zubarev gives a complete macroscopic integral convolution with thermodynamic forces, and in Ref. [27], the internal, purely operator part of this integral is deciphered through the visual geometry of the stochastic boundaries of the phase space.

There is a profound mathematical difference between the linear representation of the classical NSO method and the nonlinear representation of first-passage time thermodynamics (TFPT) developed in Ref. [27]. If the thermodynamic forces $F_k(t')$ are under the integral over time, we indeed have no right to apply the standard Laplace transform over time to the operator part separately from the forces. The convolution of the kernel and forces in Laplace space would turn into a product only if the forces $F_k$ were constants, which is incorrect in a highly nonequilibrium process (e.g., explosive boiling). This problem is solved by changing the closure scheme of the equations. The structures in Laplace space are equated based on the principle of quasi-stationarity of the local ensemble.

How thermodynamic forces enter into distribution parameters. In the book [19], the full dissipative part of the equation has the form of a time convolution $I(t)=\int_0^t K_{\mathbf{Zub}}(t-t')\cdot F_k(t')dt'$. In Ref. [27], a quasi-equilibrium operator $\rho_{rel}(t)$ is constructed not on the basis of local hydrodynamic forces $F_k(t)$, but on the basis of a fixed stochastic absorption boundary $\partial\Omega$. In his formalism, the governing thermodynamic force is the parameter γ, which is conjugated with the lifetime of the metastable state.

Forces $F_k(t')$ (e.g., liquid superheating or potential gradients) change slowly over the macroscopic time interval before boiling compared to rapid microscopic molecular fluctuations (Bogolyubov's timescale separation hypothesis). During the nucleation incubation period, macroscopic forces can be considered quasi-constant on the scale of fluctuation wandering.

In this case, the current values of the macroforces $F_k(t)$ determine the physical parameters of the lifetime distribution $\rho(T_{fpt})$. For example:

- The magnitude of superheating of the liquid determines the height of the barrier $W_{cr}(F_k)$ that enters the Kramers time $\tau_{cr}\propto\exp(W_{cr}/k_BT)$.
- The viscosity of the medium determines the non-Markovian parameter $\alpha(F_k)$.

The mathematical meaning of the Laplace map (A.3). In identity (A.3), the Laplace variable s is not the physical time of the system's macroevolution, but the internal frequency of the stochastic random walk of the phase trajectory within the domain $\Omega$. This map is performed in the so-called "frozen" local thermodynamic state (local equilibrium state). The procedure is as follows.

The current time t and the current macroforces $F_k(t)$ are fixed. Based on these forces, the spectrum of microscopic fluctuations and the corresponding first-passage time distribution $\tilde{\rho}(s| F_k)$ are calculated. For this fixed state, the Laplace transform of the memory kernel $\tilde{K}(s)$ is calculated. The time dependence of the memory kernel $K(t-t'| F_k)$ at this point is found from the inverse Laplace transform.

After this, the kernel returns to the full macroscopic equation. The full Cattaneo or Smoluchowski integro-differential equation takes the form, where the memory kernel K itself depends nonlinearly on the current state:

$$\frac{\partial J(t)}{\partial t}=-\int_0^t K\left(t-t'\middle|F_k(t')\right)\cdot F_k(t')dt'.$$

The kernel identity $\tilde{K}_{\mathbf{Zub}}(s)\equiv\tilde{K}_{\mathbf{ren}}(s)$ is applied to the operator (structural) parts of dissipation functionals within the local-analytic approximation. In Zubarev's method, macroscopic forces $F_k(t')$ are encapsulated within the parameters of the first-passage time distribution. The introduction of the Laplace transform with respect to the stochastic frequency s serves as a tool for revealing the algebraic structure of the fluctuation memory kernel under fixed macroscopic conditions. The resulting non-Markovian kernel $K\left(t-t^{'}\middle|F_k\right)$ is then substituted back into the full time-convolution integral, restoring the explicit dependence on the evolution of thermodynamic forces.

Let us carry out a step-by-step derivation of formula (21).

1. How is the probability density function $\rho(T_{fpt})$ derived from the definition of the survival function P(t)? The relationship between the probability density function of the first passage time and the macroscopic probability of survival is a fundamental bridge between the stochastic dynamics of individual trajectories and the balance equations.

Let be $P(t;\mathbf{x}_0)$ the survival function. It determines the probability that the system, starting from point $x_0$ at time t=0, is still within the metastability region $\Omega$ at time t, without ever touching the boundary $\partial\Omega$.

$$P(t;\mathrm{x}_0)=\mathrm{Prob}(t_{fpt}>t)\,.$$

By definition of the probability distribution function, a random variable $T_{fpt}$ will take a specific value in the interval from t to t+dt if the system leaves the region $\Omega$ during this particular time interval. This means that the probability density function $\rho(T_{fpt})$ is the rate of decay of the survival function over time:

$$\rho(T_{fpt};\mathbf{x}_0)=-\frac{dP(t;\mathbf{x}_0)}{dt}_{|t=T_{fpt}}\,.$$

Transition to transport equations and a memory kernel. In non-Markovian systems with memory effects, such as superheated liquids or polymer chains, the evolution of the survival function is described by an integro-differential equation with a memory kernel K(t):

$$\frac{dP(t)}{dt}=-\int_0^t K(t-t^{'})P(t^{'})dt^{'}\,.$$

We apply a one-sided Laplace transform with respect to time $t\to s$ to both parts of the equation, taking into account that at the initial moment of time the system is guaranteed to be alive $\left(P(0)\ =\ 1\right)$:

$$s\tilde{P}(s)-1=-\tilde{K}(s)\tilde{P}(s)\,.$$

From here it is easy to express the Laplace image of the survival function through the image of the memory core:

$$\tilde{P}(s)=\frac{1}{s+\tilde{K}(s)}\,.$$

Now we transform the relationship for the probability density using Laplace $\rho(t)=-dP/dt$:

$$\tilde{\rho}(s)=-\left(s\tilde{P}(s)-P(0)\right)=1-s\tilde{P}(s)\,.$$

Let's substitute the expression for $\tilde{P}(s)$ into this equality:

$$\tilde{\rho}(s)=1-\frac{s}{s+\tilde{K}(s)}=\frac{\tilde{K}(s)}{s+\tilde{K}(s)}\,.$$

Let's perform some simple algebraic transformations to express $\tilde{K}(s)$. As a result, we obtain formula (21) for the structure of the memory core:

$$\tilde{K}(s)=s\cdot\frac{\tilde{\rho}(s)}{1-\tilde{\rho}(s)}\,.$$

This shows that the macroscopic memory of a nonequilibrium environment (the core $K(t)$) is completely and unambiguously encoded in the microscopic statistics of the lifetime of its states $\tilde{\rho}(s)$.

Thus, an operator mapping of the form (A.3) is introduced within the framework of the developed formalism as a consistent physical postulate (structural ansatz). This assumption is based on Onsager's generalized fluctuation regression hypothesis and the fluctuation-dissipation theorem of the second kind, which are also based on the physical mapping principle. The assumption of this article asserts the equivalence of the macroscopic dissipation of a multiparticle system and the time statistics of the stochastic random walks of its parameters to the absorbing boundaries of phase space. This postulate functions as a closure thermodynamic correspondence principle, translating hidden information about microflow correlators into observable macroscopic kernels of non-Markovian transport. This is a functional closure of the theory, not a pure consequence of the Liouville equations.

## References

1. T. D. Donder, L'Affinitè (Gauthier-Villars, Paris, France, 1936).
2. I. Prigogine, Etude Thermodinamique des Phènomènes Irrèversibles (Desoer, Liege, Belgium, 1947).
3. L. Onsager and S. Machlup, Phys. Rev. **9**, 1505 (1953).
4. S. de Groot and P. Mazur, Nonequilibrium Thermodynamics (North Holland, Amsterdam, The Netherlands, 1962).
5. P. Glansdorff and I. Prigogine, Thermodynamic Theory of Structure, Stability, and Fluctuations (Wiley-Interscience. New York, USA, 1971).
6. C. Truesdell, Rational Thermodynamics (McGraw-Hill, New York, USA, 1985) [second enlarged edition (Springer, Berlin, Germany)] (1988)].
7. D. Jou, J. Casas-Vazquez and G. Lebon, Extended Irreversible Thermodynamics (Springer, Berlin, 1993), First edition. Second edition (1996). Third edition (2001). Fourth edition (2010).
8. G. Lebon, D. Jou, J. Casas-Vázquez, Understanding Non-equilibrium Thermodynamics, Foundations, Applications, Frontiers (Springer, Berlin, Heidelberg, 2008).
9. R. V. Velasco, and L. S. García-Colín, J. Non-Equilib. Thermodyn. **18**, 157 (1993).
10. I. Gyarmati, J. Non-Equilib. Thermodyn. **2**, 233-260 (1977).
11. M. Grmela, Phys. Rev. E, **48**:2, 919–930 (1993).
12. U. Seifert, The European Physical Journal B, **64** (3–4): 423–431 (2008); arXiv:0710.1187.
13. A. Hobson, J. Chem. Phys. **45**, 1352 (1966).
14. L. S. Garcia-Colin, Á. R. Vasconcellos and R. Luzzi, J. Non-Equilib. Thermodyn. **19**, 24 (1994).
15. R. Luzzi, Á. R. Vasconcellos and J. G. Ramos, Statistical Foundations of Irreversible Thermodynamics (Teubner-Bertelsmann Springer, Sttutgart, Germany, 2000).
16. R. Luzzi, Á. R. Vasconcellos and J. G. Ramos, La Rivista del Nuovo Cimento, **29**:2, 1–82 (2006).
17. R. Luzzi, Á. R. Vasconcellos and J. G. Ramos, et al. Theor. Math. Phys. **194**, 4-29 (2018). https://doi.org/10.1134/.
18. D. N. Zubarev, Non-equilibrium statistical thermodynamics. Plenum-Consultants Bureau, New York, USA (1974).
19. D. N. Zubarev, V. Morozov and G. Röpke, Statistical Mechanics of Non-equilibrium Processes: Basic Concepts, Kinetic Theory, (Akademie-Wiley VCH, Berlin, Germany, Vol. 1, 1996).
20. D. N. Zubarev, V. Morozov and G. Röpke, Statistical Mechanics of Non-equilibrium Processes: Relaxation and Hydrodynamic Processes, (Akademie-Wiley VCH, Berlin, Germany, Vol. 2, 1997).
21. D. N. Zubarev, in Reviews of Science and Technology: Modern Problems of Mathematics. Vol.**15**, 131-226, (in Russian) ed. by R. B. Gamkrelidze (Nauka, Moscow, 1980) [English Transl.: J. Soviet Math. **16**, 1509-1571 (1981)].
22. G. Röpke, Nonequilibrium Statistical Operator. In: Non-Equilibrium Particle Dynamics; Kim A.S., Ed.; Intech Open: London, UK (2019); ISBN 978-1-83968-079-3, doi:10.5772/intechopen.84707.
23. V. V. Ryazanov, Fortschritte der Phusik/Progress of Physics. **49**, N8-9, 885-893 (2001).
24. V. V. Ryazanov, Eur. Phys. J. B, **94**, 242 (2021). https://doi.org/10.1140/epjb/s10051-021-00246-0.
25. V. V. Ryazanov, S. G. Shpyrko, Condensed Matter Physics, **9**, 1(45), 71-80 (2006).
26. V. V. Ryazanov, European Physical Journal B. **72**, 629–639 (2009).
27 V. V. Ryazanov, Continuum Mechanics and Thermodynamics. — 2024. — Vol. 36, No. 6. — P. 1493–1513. https://doi.org/10.1007/s00161-024-01311-6.

28. First-Passage Phenomena and Their Applications, edited by R. Metzler, G. Oshanin, and S. Redner (World Scientific, Singapore, 2014) 608 p.
29. J. Masoliver, Random Processes: First-Passage and Escape (World Scientific, Singapore, 2018) 388 p.
30. V. V. Ryazanov, Journal of thermodynamics, Volume **2011**, Article ID 203203, 10 pages (2011). doi:10.1155/2011/203203.
31. V. V. Ryazanov, Journal of Thermodynamics, Vol. **2012**, Article ID 318032, 5 pages (2012). doi:10.1155/2012/318032.
32. B. B. Markiv, I. P. Omelyan, M. V. Tokarchuk, Condensed Matter Physics, **12**(4), 573–580 (2009).
33. M. C. Schwarzwälder, Non-Fourier Heat Conduction. The Maxwell-Cattaneo Equations, Master's Degree Thesis (Universitat Politècnica de Catalunya Facultat de Matemàtiques i Estadística, 2015).
34. R. Metzler, J. Klafter, Physics Reports, Volume 339, Issue 1, December, Pages 1-77 (2000).
35. E. Barkai, R. Metzler, and J. Klafter, Phys. Rev. E 61, 132 (2000).
36. U. Seifert, Stochastic Thermodynamics (Cambridge University Press, 2025).
37. U. Seifert, Reports on Progress in Physics, 75(12), 126001 (2012).
38. U. Seifert, The European Physical Journal B, 64(3), 423-431 (2005).
39. C. Jarzynski, Physical Review Letters, 78(14), 2690 (1997).
40. G. E. Crooks, *Physical Review E*, 60(3), 2721 (1999).
41. H. C. Tuckwell and F. Y. M. Wan's, Journal of Applied Probability, 21(4) (1984) DOI: 10.2307/3213688.
42. A. Raghu and I, Neri, New Journal of Physics, **27**, 124602 (2025) DOI 10.1088/1367-2630/ae249e.
43. A. Wardak, *J. Phys. A: Math. Theor.* **53**, 375001. DOI 10.1088/1751-8121/ab8b37 (2020)